\documentclass[twocolumn,tighten,astrosymb,trackchanges,times]{aastex631}
\usepackage{hyperref}
\usepackage[dvipsnames]{xcolor}

\definecolor{darkaqua}{rgb}{0.0,0.45,0.65}

\definecolor{maroon}{rgb}{0.760,0.118,0.337}

\newcommand{\Rutgers}{Department of Physics and Astronomy, Rutgers, the State University of New Jersey, Piscataway, NJ 08854, USA}

\begin{document}

\title{JWST Nebular Spectroscopy of SN\,2023qov: Circumstellar Dust Emission in a Normal Type Ia Supernova}

\author[0000-0002-9209-2787]{Colin W.~Macrie}
\affiliation{Department of Physics and Astronomy, Purdue University, 525 Northwestern Avenue, West Lafayette, IN 47907-2036, USA}
\affiliation{\Rutgers}
\email{cmacrie@purdue.edu}

\author[0000-0003-2037-4619]{Conor Larison}
\affiliation{Space Telescope Science Institute, 3700 San Martin Drive, Baltimore, MD 21218, USA}

\author[0000-0001-8023-4912]{Huei Sears}
\affiliation{\Rutgers}

\author[0000-0003-3108-1328]{Lindsey A.~Kwok}
\thanks{NASA Hubble Fellow}
\affiliation{Center for Interdisciplinary Exploration and Research in Astrophysics (CIERA), Northwestern University, Evanston, IL 60208, USA}

\author[0000-0001-8738-6011]{Saurabh W.~Jha}
\affiliation{\Rutgers}

\author[0000-0001-5694-3399]{Mi Dai}
\affiliation{Department of Physics and Astronomy and PITT PACC, University of Pittsburgh, Pittsburgh, PA 15260, USA}

\author[0000-0001-5975-290X]{Joel~Johansson}
\affiliation{Department of Physics, Oskar Klein Centre, Stockholm University, SE-106 91, Stockholm, Sweden}

\author[0000-0002-9388-2932]{St\'ephane~Blondin}
\affiliation{European Southern Observatory, Karl-Schwarzschild-Str. 2, 85748, Garching, Germany}
\affiliation{Aix-Marseille Université, CNRS, CNES, LAM, Marseille, France}

% \author[]{}
% \affiliation{}

\author[0000-0002-1895-6639]{Moira Andrews}
\affiliation{Las Cumbres Observatory, 6740 Cortona Dr. Suite 102, Goleta, CA 93117, USA}
\affiliation{Department of Physics, University of California, Santa Barbara, Santa Barbara, CA 93106, USA}

\author[0000-0002-4449-9152]{K.~Auchettl}
\affiliation{School of Physics, The University of Melbourne, Parkville, VIC, Australia}
\affiliation{Department of Astronomy and Astrophysics, University of California, Santa Cruz, CA 95064, USA}

\author[0000-0003-3494-343X]{Carles Badenes}
\affiliation{Department of Physics and Astronomy and PITT PACC, University of Pittsburgh, Pittsburgh, PA 15260, USA}

\author[0000-0003-4769-4794]{Barnab\'as~Barna}
\affiliation{Department of Experimental Physics, Institute of Physics, University of Szeged, D{\'o}m t{\'e}r 9, 6720 Szeged, Hungary}
\affiliation{HUN-REN-SZTE Stellar Astrophysics Research Group, 6500 Baja, Szegedi \'ut, Kt. 766, Hungary}

\author[0000-0002-4924-444X]{K.~Azalee~Bostroem}
\affiliation{Steward Observatory, University of Arizona, 933 North Cherry Avenue, Tucson, AZ 85721-0065, USA}

\author[0000-0001-5955-2502]{Thomas~G.~Brink}
\affiliation{Department of Astronomy, University of California, Berkeley, CA  94720-3411}

\author[0000-0002-5680-4660]{Kyle~W.~Davis}
\affiliation{Department of Astronomy and Astrophysics, University of California, Santa Cruz, CA 95064, USA}

\author[0000-0003-4914-5625]{Joseph~R.~Farah}
\affiliation{Las Cumbres Observatory, 6740 Cortona Dr. Suite 102, Goleta, CA 93117, USA}
\affiliation{Department of Physics, University of California, Santa Barbara, Santa Barbara, CA 93106, USA}

\author[0000-0003-3460-0103]{Alexei~V.~Filippenko}
\affiliation{Department of Astronomy, University of California, Berkeley, CA 94720-3411}

\author[0000-0003-2238-1572]{Ori~D.~Fox}
\affiliation{Space Telescope Science Institute, 3700 San Martin Drive, Baltimore, MD 21218, USA}

\author[0000-0002-4391-6137]{Or~Graur}
\affiliation{Institute of Cosmology and Gravitation, University of Portsmouth, Portsmouth, PO1 3FX, UK}
\affiliation{Department of Astrophysics, American Museum of Natural History, Central Park West and 79th Street, New York NY 10024-5192, USA}

\author[0000-0002-3841-380X]{Saarah~Hall}
\affiliation{Department of Physics and Astronomy, Northwestern University, 2145 Sheridan Road, Evanston, IL 60208, USA}
\affiliation{Center for Interdisciplinary Exploration and Research in Astrophysics (CIERA), Northwestern University, Evanston, IL 60208, USA}

\author[0000-0003-4253-656X]{D.~Andrew~Howell}
\affiliation{Las Cumbres Observatory, 6740 Cortona Dr. Suite 102, Goleta, CA 93117, USA}
\affiliation{Department of Physics, University of California, Santa Barbara, Santa Barbara, CA 93106, USA}

\author[0000-0002-0832-2974]{Griffin Hosseinzadeh}
\affiliation{Department of Astronomy \& Astrophysics, University of California, San Diego, 9500 Gilman Drive, MC 0424, La Jolla, CA 92093-0424, USA}

\author[0000-0001-8005-4030]{Anders Jerkstrand}
\affiliation{Department of Physics, Oskar Klein Centre, Stockholm University, SE-106 91, Stockholm, Sweden}

\author[0000-0002-8770-6764]{R\'eka K\"onyves-T\'oth}
\affiliation{HUN-REN Research Centre for Astronomy and Earth Sciences, Konkoly Observatory, Konkoly Th. M. {\'u}t 15-17., 1121 Budapest, Hungary}
\affiliation{CSFK, MTA Centre of Excellence, Konkoly Thege Mikl{\'o}s {\'u}t 15-17, 1121 Budapest, Hungary}

\author[0000-0003-1731-0497]{C.~Lidman}
\affiliation{The Research School of Astronomy and Astrophysics \& Centre for Gravitational Astrophysics,\\
The Australian National University, Canberra, ACT 2611, Australia}

\author[0000-0003-2611-7269]{Keiichi Maeda}
\affiliation{Department of Astronomy, Kyoto University, Kitashirakawa-Oiwake-cho, Sakyo-ku, Kyoto 606-8502, Japan}

\author[0000-0002-9770-3508]{Kate~Maguire}
\affiliation{School of Physics, Trinity College Dublin, College Green, Dublin 2, Ireland}

\author[0009-0006-4963-3206]{Bailey Martin}
\affiliation{The Research School of Astronomy and Astrophysics \& Centre for Gravitational Astrophysics,\\
The Australian National University, Canberra, ACT 2611, Australia}

\author[0000-0001-9515-478X]{Adam~Miller}
\affiliation{Department of Physics and Astronomy, Northwestern University, 2145 Sheridan Road, Evanston, IL 60208, USA}
\affiliation{Center for Interdisciplinary Exploration and Research in Astrophysics (CIERA), Northwestern University, Evanston, IL 60208, USA}
\affiliation{NSF-Simons AI Institute for the Sky (SkAI), 172 E. Chestnut St., Chicago, IL 60611, USA}

\author[0000-0001-9570-0584]{Megan~Newsome}
\affiliation{Department of Physics and Astronomy, University of Texas Austin, 2515 Speedway, Austin, TX 78712}
\affiliation{Las Cumbres Observatory, 6740 Cortona Dr. Suite 102, Goleta, CA 93117, USA}
\affiliation{Department of Physics, University of California, Santa Barbara, Santa Barbara, CA 93106, USA}

\author[0000-0003-0209-9246]{Estefania Padilla Gonzalez}
\affiliation{Physics and Astronomy Department, Johns Hopkins University, Baltimore, MD 21218, USA}

\author[0000-0002-1633-6495]{Abigail~Polin}
\affiliation{Department of Physics and Astronomy, Purdue University, 525 Northwestern Avenue, West Lafayette, IN 47907-2036, USA}

\author[0000-0002-4410-5387]{Armin Rest}
\affiliation{Space Telescope Science Institute, 3700 San Martin Drive, Baltimore, MD 21218, USA}
\affiliation{Physics and Astronomy Department, Johns Hopkins University, Baltimore, MD 21218, USA}

\author[0009-0006-0764-8856]{Zoe~A.~Rosenberg}
\affiliation{Center for Computational Relativity and Gravitation, Rochester Institute of Technology, Rochester, New York 14623, USA}

\author[0000-0003-4102-380X]{David~J.~Sand}
\affiliation{Steward Observatory, University of Arizona, 933 North Cherry Avenue, Tucson, AZ 85721-0065, USA}

\author[0009-0002-5096-1689]{Michaela~Schwab}
\affiliation{Department of Astronomy, University of Virginia, 530 McCormick Road, Charlottesville, VA 22904, USA}

\author[0000-0003-2445-3891]{Matthew~R.~Siebert}
\affiliation{Space Telescope Science Institute, 3700 San Martin Drive, Baltimore, MD 21218, USA}

\author[0000-0001-6706-2749]{Mridweeka Singh}
\affiliation{Indian Institute of Astrophysics, Koramangala 2nd Block, Bangalore 560034, India}

%\author[0000-0003-2445-3891]{Matthew~R.~Siebert}
%\affiliation{Space Telescope Science Institute, 3700 San Martin Drive, Baltimore, MD 21218, USA}

\author[0000-0003-4610-1117]{Tam\'as Szalai}
\affiliation{Department of Experimental Physics, Institute of Physics, University of Szeged, D{\'o}m t{\'e}r 9, 6720 Szeged, Hungary}
\affiliation{MTA-ELTE Lend\"ulet "Momentum" Milky Way Research Group, Szent Imre H. st. 112, 9700 Szombathely, Hungary}

\author[0000-0001-7380-3144]{Tea Temim}
\affiliation{Department of Astrophysical Sciences, Princeton University, Princeton, NJ 08544, USA}

\author[0000-0001-9834-3439]{Jacco~H.~Terwel}
\affiliation{School of Physics, Trinity College Dublin, College Green, Dublin 2, Ireland}

\author[0000-0002-4283-5159]{Brad~E.~Tucker}
\affiliation{$^{}$Mt Stromlo Observatory, The Research School of Astronomy and Astrophysics, Australian National University, ACT 2611, Australia\\}
\affiliation{$^{}$Australian National Centre for the Public Awareness of Science, Australian National University, ACT 2601, Australia\\}

\author[0000-0001-8764-7832]{J\'ozsef Vink\'o}
\affiliation{HUN-REN Research Centre for Astronomy and Earth Sciences, Konkoly Observatory, Konkoly Th. M. {\'u}t 15-17., 1121 Budapest, Hungary}
\affiliation{ELTE E\"otv\"os Lor\'and University, Institute of Physics and Astronomy, P\'azm\'any P\'eter s\'et\'any 1, Budapest, Hungary}
\affiliation{Department of Experimental Physics, Institute of Physics, University of Szeged, D{\'o}m t{\'e}r 9, 6720 Szeged, Hungary}

\author[0000-0002-1094-3817]{Lingzhi Wang}
\affiliation{College of Science, Hainan Tropical Ocean University, Sanya, 572022, China}
\affiliation{Chinese Academy of Sciences South America Center for Astronomy (CASSACA), National Astronomical Observatories, CAS, Beijing 100101, China}

\author[0000-0002-7334-2357]{Xiaofeng~Wang}
\affiliation{Department of Physics, Tsinghua University, Beijing, 100084, China}

\author[0000-0002-2636-6508]{WeiKang~Zheng}
\affiliation{Department of Astronomy, University of California, Berkeley, CA 94720-3411}

% \author{Others}

%% Note that the \and command from previous versions of AASTeX is now
%% depreciated in this version as it is no longer necessary. AASTeX 
%% automatically takes care of all commas and "and"s between authors names.

%% AASTeX 6.31 has the new \collaboration and \nocollaboration commands to
%% provide the collaboration status of a group of authors. These commands 
%% can be used either before or after the list of corresponding authors. The
%% argument for \collaboration is the collaboration identifier. Authors are
%% encouraged to surround collaboration identifiers with ()s. The 
%% \nocollaboration command takes no argument and exists to indicate that
%% the nearby authors are not part of surrounding collaborations.

%% Mark off the abstract in the ``abstract'' environment. 
\begin{abstract}
We present panchromatic observations of the Type Ia supernova (SN~Ia) 2023qov, ranging from $\sim$2 weeks before to $\sim$1 year after maximum light. \textit{JWST} near- and mid-infrared spectra at $+$276 and $+$363~days show $\sim$400 K dust emission that cools by $\sim$75~K between epochs, the first unambiguous spectroscopic detection of dust emission in a normal SN~Ia. We find that the emission is well described by models of carbonaceous dust placed within $\sim$1 light year of the SN, with a dust mass of $\sim$$10^{-4}$\,M$_{\odot}$. We do not see evidence of active dust creation, suggesting an infrared light echo by pre-existing circumstellar dust as the likely source of the emission. The \textit{JWST} nebular line profiles suggest asymmetric, stratified ejecta, similar to  other normal SNe~Ia, though a slight double-horn structure in the argon lines indicate a toroidal enhancement. SN\,2023qov exhibits a slightly red, fast-declining early light curve ($\Delta m_{15}(B) = 1.47 \pm 0.05$\,mag), from which we determine a $^{56}$Ni mass of $M_{56} = 0.21 \pm 0.04$\,M$_{\odot}$, and a distance of $d = 36.0 \pm 1.8$\,Mpc to the SN and its host, NGC 7029. The presence of circumstellar dust around a normal SN~Ia has implications both for the progenitor system, suggesting a less than pristine environment, and for cosmology, as a possible source of reddening and extinction that can affect distance inferences.
\end{abstract}


\section{Introduction} \label{sec:intro}
% intro to snIa 
% and with lower temperature estimates of $\sim$355 and $\sim$310 for the favored model

Type Ia supernovae (SNe~Ia) are the thermonuclear explosions of carbon-oxygen white dwarfs in binary systems \citep{Hoyle1960}. Though they are used extensively as distance indicators in cosmological studies \citep[e.g.,][]{phillips_absolute_1993,tripp_two-parameter_1998}, including in the discovery of the accelerating expansion of the Universe \citep{Riess_expansion_1998,perlmutter_measurements_1999}, their progenitor systems and explosion mechanisms are not concretely understood. In current literature, there are two primary progenitor models: a single-degenerate (SD) scenario where a white dwarf interacts with a main-sequence, red-giant, or helium-star companion, or a double-degenerate (DD) system comprised of a pair of white dwarfs. Further, multiple explosion mechanisms are possible, with models that span a range of deflagration and detonation scenarios \citep[for recent reviews, see][]{BLONDIN_review_2026,Ruiter:2025,Liu_review_2023}.

Contemporary observations have brought a wealth of SN~Ia data to address these questions \citep{jha_observational_2019}. High-cadence surveys, including both wide field \citep{holoien_asas-sn_2017,tonry_atlas_2018,bellm_zwicky_2019} and targeted programs \citep{tartaglia_dlt40_2018,itagaki_transient_2020}, enable the early discovery of nearby SNe. Comprehensive datasets of these objects now often include ultraviolet (UV) through near-infrared (NIR) photometry and spectroscopy \citep[e.g.,][]{Brown:2014,silverman_berkeley_2012,Morrell:2024,Tinyanont:2024}. A new addition is late-time nebular NIR and mid-infrared (MIR) spectroscopy with \textit{JWST} \citep[e.g.,][]{Kwok_2021aefxJWST_2023,Kwok_22pul_2024, Kwok_24pxl_2025, Kwok_2022aaiq24gy_2025,Dercacy_2021aefx_2023,Derkacy_22xkq_2024,Ashall_21aefxJWST_2024,Siebert:2024}. Extensive studies of individual, nearby SNe~Ia provide our best avenue for understanding their astrophysical origins.

Spectral time series offers an in-depth, evolving view of an SN by revealing continuously deeper layers of the SN ejecta as the photosphere recedes into the innermost, lower-velocity regions. At late times ($\gtrsim$150 days post-maximum; see \citealt{fransson_reconciling_2015}), the SN~Ia ejecta become increasingly optically thin, allowing us to see through the SN and directly observe the velocity structure of the elements synthesized in the explosion (e.g., \citealt{Jerkstrand_lineprofiles_2015}).
Importantly, with late-time \textit{JWST} observations, we can observe nebular lines of intermediate-mass elements (IMEs; e.g., Si, Ar, Ca) and iron-group elements (IGEs; e.g., Fe, Co, Ni) whose abundances are sensitive to the density and temperature structure of the progenitor.
\textit{JWST} observations of several normal SNe~Ia (e.g., SN\,2021aefx, \citealt{Kwok_2021aefxJWST_2023,Dercacy_2021aefx_2023,Ashall_21aefxJWST_2024}; SN\,2022aaiq and SN\,2024gy, \citealt{Kwok_2022aaiq24gy_2025}) have shown strong central emission from stable nickel, indicative of high-density burning \citep{Blondin_nickel_2022} --- perhaps implying a near-Chandrasekhar-mass WD progenitor --- as well as IME line profiles that provide evidence for stratified ejecta, a hallmark of detonations (e.g., delayed detonation or double detonation).

% supernova environments
With a growing sample of \textit{JWST} observations of SNe~Ia, we can begin to connect these nebular, NIR+MIR SN properties to their host-galaxy environments. SN\,2022aaiq \citep{Kwok_2022aaiq24gy_2025} and SN\,2023qov occurred in quiescent host galaxies, while SN\,2021aefx \citep{Kwok_2021aefxJWST_2023, Dercacy_2021aefx_2023, Ashall_21aefxJWST_2024}, SN\,2022xkq \citep{Derkacy_22xkq_2024}, and SN\,2024gy \citep{Kwok_2022aaiq24gy_2025} are in star-forming hosts. Evidence shows that SN~Ia peak luminosity and light-curve evolution are environment-dependent \citep{branch_statistical_1996,hamuy_absolute_1996, hamuy_search_2000,sullivan_rates_2006,sullivan_dependence_2010,rigault_evidence_2013,Wang_2013_population, Wang_2019}. Samples of normal SNe~Ia exhibit a bimodal distribution of the decline rate \citep{Wojtak_bimodal_2023,Larison_environments_2024,Senzel_hosts_2024,padilla_gonzalez_2025}, with hosts of the faster declining SNe occurring being older galaxies with low star-formation rates \citep{rigault_strong_2020}. The slower-evolving SNe~Ia usually occur in star-forming galaxies. Conversely, the peak colors of SNe~Ia appear unimodal, not depending on the host environment \citep{Wojtak_bimodal_2023,Larison_environments_2024,Senzel_hosts_2024}. Different progenitors could explain the bimodality in decline rate, as older stellar populations may support different progenitor systems. If progenitor-system characteristics were to explain this bimodality, differences between the two populations could be revealed in nebular NIR+MIR emission. 

SNe~Ia are the main contributors to the chemical enrichment of the Universe with IGEs \citep{Nomoto:2013}. However, when it comes to dust, there has been little observational evidence to suggest that normal SNe~Ia contribute to the formation of interstellar dust \citep{Gomez_dust_2007,gerardy_signatures_2007}. Thermonuclear SNe are expected to be hostile to pre-existing dust, destroying any created by the progenitor system during the explosion \citep{Morgan_Dust_2003}. 

Some work has already been done to constrain the possibility of circumstellar material (CSM) around normal SNe~Ia with large samples in the NIR \citep{Maeda_dustHJK_2015}, MIR \citep{Johansson_dust_2013}, and radio \citep{chomiuk_deep_2016}. These have revealed that many observations of normal SNe~Ia have not yet ruled out CSM presence altogether, and have given upper limits of pre-explosion mass loss ranging from $10^{-9}$ to $10^{-4}$\,M$_{\odot}$\,yr$^{-1}$. Further, recent evidence has shown that peculiar SN~Ia environments are not as clean as previously thought. Late-time MIR emission from the peculiar SN~Ia, SN\,2018evt \citep{Wang_CSMDust_2024}, revealed $\sim 10^{-2}$\,M$_{\odot}$ of dust formation at $\gtrsim$1000~days post-explosion. This dust creation is thought to be powered by the interaction between the SN ejecta and hydrogen-rich CSM, similar to the Type Ia-CSM SN 2005gj \citep{fox_05gj_2013}. 

Recently, evidence for dust has also arisen in peculiar SNe~Ia not normally associated with rich CSM, with \citet{Mo_CSM_2025} reporting MIR rebrightenings in five SNe~Ia of the peculiar ``Iax'' (SNe~Iax) \citep{foley_type_2013}, ``91T-like,'' \citep{Filippenko_91T_1992a,phillips_absolute_1993} and ``03fg-like'' \citep{howell_type_2006} subtypes. A similar rebrightening was observed in the Type Iax SN 2014dt \citep{fox_excess_2016}. This, combined with improved modeling \citep[e.g.,][]{Kumar_Iax_2026}, has shown that low-luminosity thermonuclear SNe, such as SNe~Iax, are strong candidates to contribute to interstellar dust populations.

Despite these advances, the potential origin of dust in SNe~Ia is not well understood. \textit{JWST} MIR spectroscopy of the peculiar 03fg-like or ``super-Chandrasekhar'' SN\,2022pul revealed a dust continuum well fit by a blackbody at $T\approx500$\,K \citep{Siebert:2024}. Supernova remnants (SNRs) Tycho and Kepler (both predicted to have been normal SNe~Ia) show evidence of dust emission, though their dust masses may be consistent with that which would be collected in the interstellar medium (ISM) as the SN ejecta expand \citep{GomezClark_Remnantdust_2012}. \cite{Court_SNRCSM_2026} sampled Type Ia SNRs and determined that nearly half show bulk ejecta profiles consistent with CSM interaction beyond what is expected from the ISM alone. Whether normal SNe~Ia can produce or heat detectable dust on $\sim$100\,d timescales, corresponding to nebular epochs, remains an open question that \textit{JWST} can address.
% Increasing evidence has thus shown that SNe~Ia environments are not as clean as previously thought, and the characterization of the molecular dust found in them is paramount to understanding their progenitors and evolution. 

Here we present two epochs of contemporaneous optical + NIR + MIR nebular spectroscopy of the normal Type~Ia SN\,2023qov obtained with ground-based facilities and \textit{JWST}. We report a clear detection of a cooling dust continuum that is consistent with emission from warm, carbonaceous circumstellar dust. Moreover, we present early-time light-curve modeling and velocity measurements to investigate progenitor and explosion scenarios, comparing against other well-studied regular SNe~Ia. SN\,2023qov is the fastest-declining normal SN~Ia of the published \textit{JWST} sample so far and thus provides the opportunity to test the nebular properties of fast-declining but otherwise normal SNe~Ia. \autoref{sec:Obs} presents our ground-based and \textit{JWST} observations of SN\,2023qov. In \autoref{sec:Photosepheric_phase}, we model the early light curves, compute a luminosity distance, and compare color and \ion{Si}{2} velocities to other SNe~Ia. In \autoref{sec:NebularLineProfiles}, we identify and fit the emission lines and dust continuum in the nebular \textit{JWST} spectra. Our results and the implications of our characterization of SN\,2023qov are discussed in \autoref{sec:Discussion}.

% spectra

\begin{figure}
    \centering
    \includegraphics[width=1\linewidth]{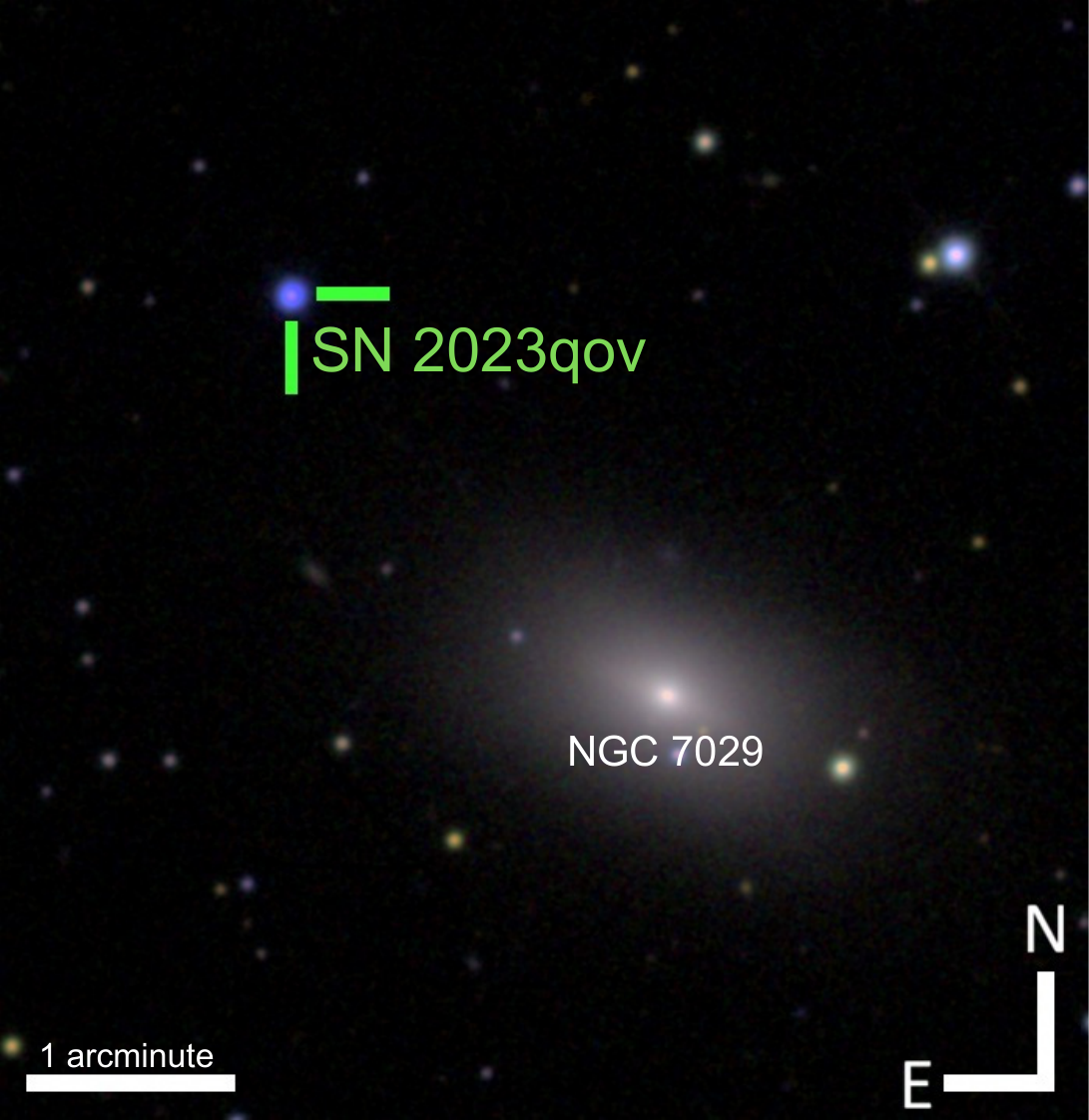}
\caption{Las Cumbres Observatory \textit{gri} image of SN\,2023qov and its elliptical host galaxy NGC\,7029 four days before peak brightness. Based on our computed distance to SN\,2023qov and the angular separation, its projected distance is 28\,kpc from the host center.}
    \label{fig:image}
\end{figure}

\section{Observations} \label{sec:Obs}

\subsection{Discovery \& Classification}

SN\,2023qov (\autoref{fig:image}) was discovered by the Asteroid Terrestrial-impact Last Alert System (ATLAS) sky survey on 2023-08-23 04:46:25 UTC at J2000 coordinates of $\alpha = 21^{\rm hr}12^{\rm m}02.036^{\rm s}$ and $\delta = -49^\circ15'18.07''$ \citep{tonry_atlas_2018,Aamer_ATLASdiscovery_2023}. It was found at a brightness of $17.51 \pm 0.03$\,mag in the ATLAS \textit{c} band, with a last nondetection four nights prior (2023-08-19) at a lower limit of 19.28\,mag. The supernova has an angular separation of 142.2$\arcsec$ from the center of its host galaxy NGC 7029 (redshift $z= 0.00947$; \citealt{Ogando_NGC7029_2008}); based on our calculated distance (\autoref{subsec:distance}), this corresponds to a projected physical separation of $\sim$28\,kpc.  \cite{Gonzalez_Classification_2023} classified SN\,2023qov to be a normal SN~Ia using the Supernova Identification (SNID) classification software \citep{Blondin_SNID_2007}.

\subsection{Photometry}

\begin{figure*}
    \centering
    \includegraphics[width=.8\linewidth]{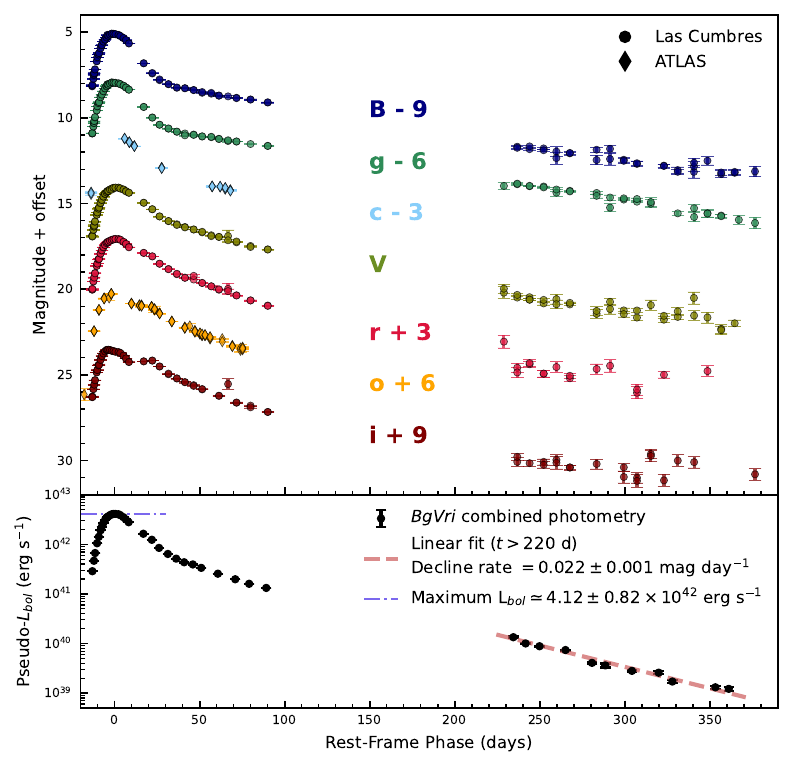}
    \caption{\textit{Top:} Bessel \textit{BV}- and SDSS \textit{gri}-band photometry of SN\,2023qov from Las Cumbres Observatory and ATLAS \textit{oc}-band photometry. The light curve is shown relative to $B$-band maximum light. Apparent magnitudes are corrected for line-of-sight Milky Way dust extinction. \textit{Bottom:} the pseudobolometric light curve, combining the \textit{BgVri} bands with the built-in \texttt{SNooPy} bolometric light curve routine (see \autoref{subsec:Lightcurves}).}
    \label{fig:phot}
\end{figure*}

Following the discovery of SN\,2023qov, extensive multiwavelength, photometric follow-up observations were triggered by the Las Cumbres Observatory network via the \textit{Global Supernova Project} in the \textit{BgVri} filters, reduced via the LCOGTsnpipe pipeline\footnote{\url{https://github.com/LCOGT/lcogtsnpipe}}. Additional imaging by ATLAS contributed a well-sampled light curve in the \textit{oc} bands (\autoref{fig:phot}).

%the Distance Less Than 40 (DLT40) Mpc survey \citep{tartaglia_dlt40_2018} and 
\subsection{Ground-Based Spectroscopy}

\begin{figure*}
    \centering
    \includegraphics[width=.73\textwidth]{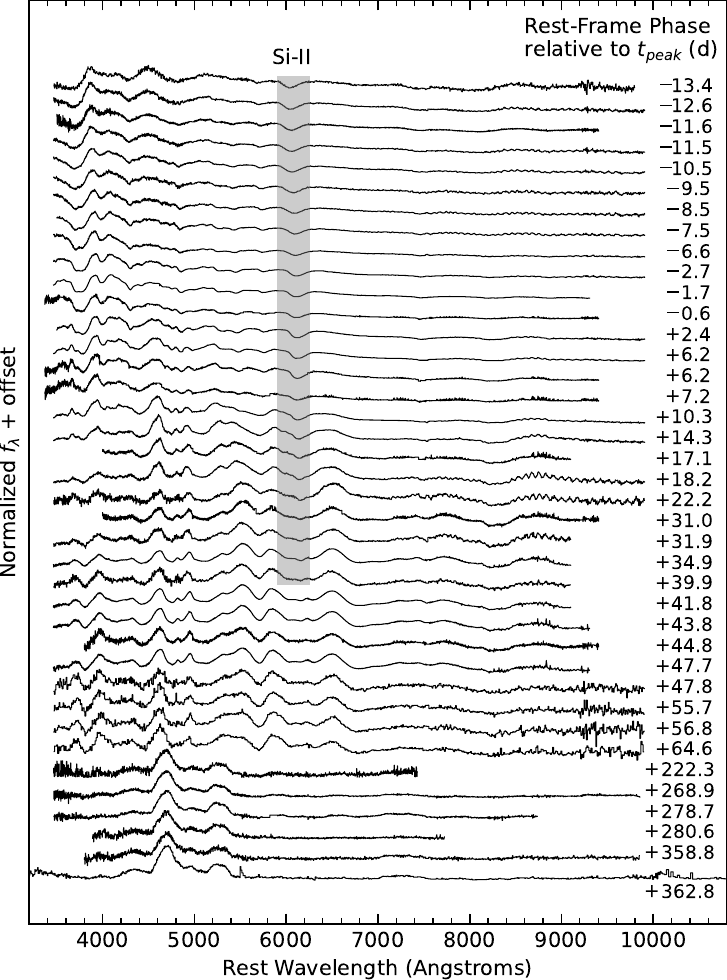}
    \caption{Spectral time series of SN\,2023qov, with scaled flux density and wavelength in the rest frame. The spectra are labeled by rest-frame phase relative to the \textit{B}-band peak. The \ion{Si}{2} $\mathrm{\lambda}$6355 absorption feature is highlighted in the early spectra. The full list of observations is presented in \autoref{tab:Spectra}.}
    \label{fig:spectra}
\end{figure*}

21 spectra were obtained with the FLOYDS specrometer \citep{Sand_FLOYDS_2011} on Las Cumbres Observatory’s Faulkes Telescope South (FTS) at a high cadence as part of optical follow-up observations with the \textit{Global Supernova Project}. They were reduced using the FLOYDS reduction pipeline \citep{Valenti_FLOYDS_2014}. The earliest spectrum taken 13.4~days pre-maximum was used to classify the object \citep{Gonzalez_Classification_2023}.

Integral field spectroscopic observations of 2023qov, covering seven epochs between 2023-08-25 and 2023-10-21 (UTC dates are used throughout this paper), were obtained with the Wide Field Spectrograph (WiFeS; \citealt{Dopita_WIFES_2007,Dopita_WIFES_2010}) on the ANU 2.3m telescope. The observations were done with the B3000 and R3000 gratings in the blue and red arms, respectively, and the X570 dichroic. The data were taken in the Nod-and-Shuffle mode. The data were processed with version 2.0.0 of the PyWiFeS data reduction pipeline and used standard calibrations to remove instrumental effects and telluric features. The SN flux was then extracted from red and blue data cubes, and the resulting one-dimensional spectra from the two arms were spliced together to form a spectrum covering the 3400\,\AA\,to 9500\,\AA\,wavelength range at a spectral resolution of about 3000.

We obtained seven spectra of SN\,2023qov with the Robert Stobie Spectrograph \citep[RSS;][]{Smith_RSS_2006} on the Southern African Large Telescope (SALT), reduced via a custom Pyraf-based reduction pipeline built with standard Pyraf \citep{pyraf_2012} spectral reduction routines and the \texttt{PySALT} package \citep{Crawford_pysalt_2010}. The reduction included the removal of the host-galaxy lines and continuum, cosmic rays, and telluric absorption. These data ranged from 1.7~days pre-maximum (2023-09-03) to 281.3~days post-maximum (2024-06-14).

On 2024-06-03 and 2024-09-02, we acquired spectra with the Low Resolution Imaging Spectrometer (LRIS; \citealt{oke_keck_1995}) on the 10\,m Keck I telescope at the W. M. Keck Observatory.  These observations utilized the $1''$ slit, the D560 dichroic, the 600/4000 grism, and the 400/8500 grating.  This instrument configuration produced a combined wavelength range of $\sim 3200$--10,200\,\AA\ and a spectral resolving power of $R = \lambda/\Delta\lambda \approx 900$.  LRIS has an atmospheric dispersion corrector; nevertheless, to further minimize slit losses caused by atmospheric dispersion \citep{filippenko_importance_1982}, the slit was oriented at or near the parallactic angle. The LRIS spectra were reduced with the LPipe data-reduction pipeline \citep{Perley_LRIS_2019}.

Another Keck/LRIS spectrum was obtained on 2024-06-13. To reduce it, we used the {\tt UCSC Spectral Pipeline}\footnote{\url{https://github.com/msiebert1/UCSC\_spectral\_pipeline}} \citep{Siebert20}, a custom data-reduction pipeline based on procedures outlined by \citet{Foley03}, \citet{Silverman2012}, and references therein.  The two-dimensional (2D) spectra were bias-corrected, flat-field corrected, adjusted for varying gains across different chips and amplifiers, and trimmed. The spectra were wavelength-calibrated using internal comparison-lamp spectra with linear shifts applied by cross-correlating the observed night-sky lines in each spectrum to a master night-sky spectrum. We combine the sides by scaling one spectrum to match the flux of the other in the overlap region and use their error spectra to correctly weight the spectra when combining.  More details of this process are discussed elsewhere \citep{Foley03, Silverman2012, Siebert20}. 

  % One-dimensional (1D) spectra were extracted using the optimal algorithm \citep{Horne86}.  
% Flux calibration and telluric correction were performed using standard stars at a similar airmass to that of the science exposures.

A total of seven VLT+XSHOOTER spectra were obtained between 2024-09-06 and 2024-09-09, each with a total integration time of 2400\,s split into a $4\times 600$\,s ABBA nodding cycle. We used a $2\times 2$ spectral and spatial binning (slow readout mode) for the UVB and VIS CCDs. To match the requested $1''$ seeing, we used slits of $1''$, $0.9''$, and $0.9''$ for the UVB, VIS, and NIR arms, respectively. The spectra were reduced using the {\tt XSHOOTER} pipeline v3.6.8 via the EsoReflex environment \citep{Freudling_VLT_2013}. The final spectrum corresponds to a median-combined version of the seven individual spectra. The full ground-based spectral time series is shown in \autoref{fig:spectra} and listed in \autoref{tab:Spectra}.

%Colin: will need descriptions of keck, VLT reductions from those respective contributors

\begin{table}[]
    \centering
    \begin{tabular}{c|r|r|c}
        \hline
        \hline
        Date & MJD & Phase & Telescope/Instrument \\
         (UTC) & ~ & (days) & \\
         \hline
        2023-08-23 & 60179.7 & $-$13.4 & FTS/FLOYDS\\
        2023-08-24 & 60180.5 & $-$12.6 & FTS/FLOYDS\\
        2023-08-25 & 60181.5 & $-$11.5 & ANU 2.3\,m/WiFeS\\
        2023-08-25 & 60181.6 & $-$11.5 & FTS/FLOYDS\\
        2023-08-26 & 60182.6 & $-$10.5 & FTS/FLOYDS\\
        2023-08-27 & 60183.6 & $-$9.5 & FTS/FLOYDS\\
        2023-08-28 & 60184.6 & $-$8.5 & FTS/FLOYDS\\
        2023-08-29 & 60185.6 & $-$7.5 & FTS/FLOYDS\\
        2023-08-30 & 60186.5 & $-$6.6 & FTS/FLOYDS\\
        2023-09-03 & 60190.4 & $-$2.7 & FTS/FLOYDS\\
        2023-09-03 & 60191.5 & $-$1.7 & SALT/RSS\\
        2023-09-05 & 60192.6 & $-$0.6 & ANU 2.3\,m/WiFeS\\
        2023-09-08 & 60195.6 & +2.4 & FTS/FLOYDS\\
        2023-09-12 & 60199.4 & +6.2 & FTS/FLOYDS\\
        2023-09-12 & 60199.5 & +6.2 & ANU 2.3\,m/WiFeS\\
        2023-09-13 & 60200.5 & +7.2 & ANU 2.3\,m/WiFeS\\
        2023-09-16 & 60203.5 & +10.3 & FTS/FLOYDS\\
        2023-09-20 & 60207.6 & +14.3 & FTS/FLOYDS\\
        2023-09-23 & 60210.4 & +17.1 & ANU 2.3\,m/WiFeS\\
        2023-09-24 & 60211.5 & +18.2 & FTS/FLOYDS\\
        2023-09-28 & 60215.6 & +22.2 & FTS/FLOYDS\\
        2023-10-07 & 60224.4 & +31.0 & ANU 2.3\,m/WiFeS\\
        2023-10-08 & 60225.4 & +31.9 & FTS/FLOYDS\\
        2023-10-10 & 60228.4 & +34.9 & SALT/RSS\\
        2023-10-16 & 60233.4 & +39.9 & FTS/FLOYDS\\
        2023-10-17 & 60235.4 & +41.8 & SALT/RSS\\
        2023-10-19 & 60237.4 & +43.8 & SALT/RSS\\
        2023-10-21 & 60238.4 & +44.8 & ANU 2.3\,m/WiFeS\\
        2023-10-23 & 60241.4 & +47.7 & SALT/RSS\\
        2023-10-24 & 60241.4 & +47.8 & FTS/FLOYDS\\
        2023-11-01 & 60249.4 & +55.7 & FTS/FLOYDS\\
        2023-11-02 & 60250.5 & +56.8 & FTS/FLOYDS\\
        2023-11-10 & 60258.4 & +64.6 & FTS/FLOYDS\\
        2024-04-17 & 60417.6 & +222.3 & SALT/RSS\\
        2024-06-03 & 60464.6 & +268.9 & Keck/LRIS\\
        2024-06-10 & 60471.4 & +275.6 & JWST/NIRSpec+MIRI \\
        2024-06-13 & 60474.6 & +278.7 & Keck/LRIS \\
        2024-06-14 & 60476.5 & +280.6 & SALT/RSS\\
        2024-09-02 & 60556.0 & +358.8 & Keck/LRIS \\
        2024-09-05 & 60559.4 & +362.7 & JWST/NIRSpec+MIRI\\
        2024-09-08 & 60561.0 & +362.8 & VLT/XShooter\\
         \hline
    \end{tabular}
    \caption{Log of spectroscopic observations. Phase in the rest frame, relative to the time of peak $B$-band brightness.}
    \label{tab:Spectra}
\end{table}

\subsection{\textit{JWST} Observations}

\begin{figure*}
    \centering
    \includegraphics[width=1\linewidth]{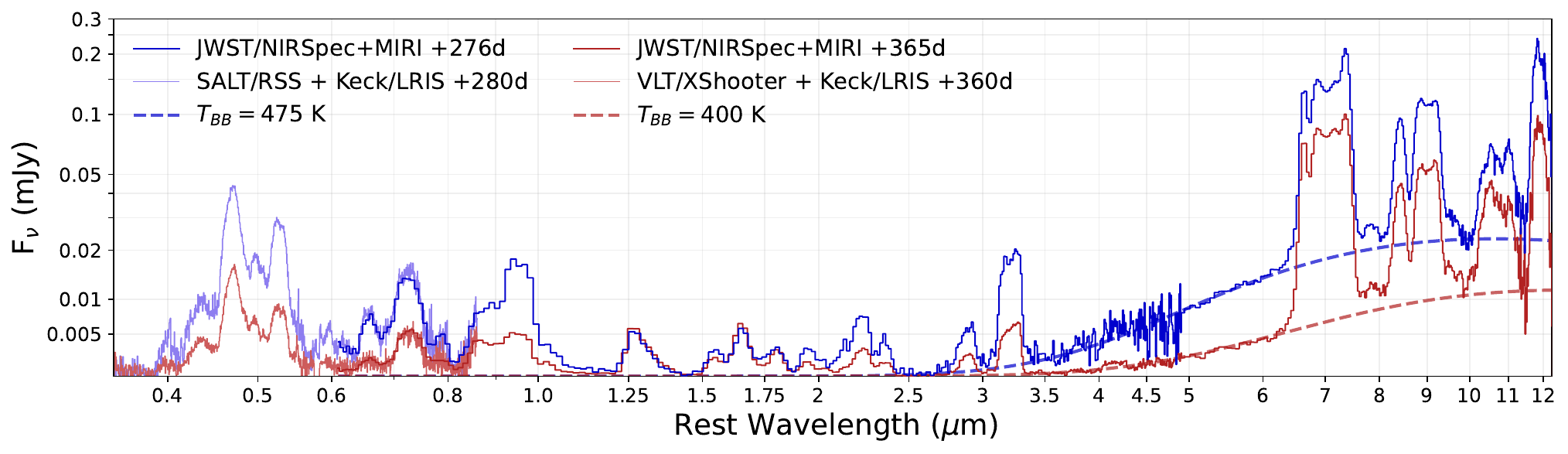}
    \caption{Optical and NIR/MIR spectra of SN\,2023qov taken at $+$276 (brighter spectra, shades of blue) and $+$363 (fainter spectra, shades of red) days since peak brightness. Dashed lines show a fading blackbody at $T = 475$\,K (blue) and $T = 400$\,K (red). These blackbody fits are discussed in \autoref{sec:NebularLineProfiles}.}
    \label{fig:JWSTspec_e1_e2}
\end{figure*}

We present NIR+MIR \textit{JWST} spectra of SN\,2023qov at $+$276 and $+$363 days in \autoref{fig:JWSTspec_e1_e2}. Observations were obtained with the Near Infrared Spectrograph (NIRSpec) in the Fixed Slit Spectroscopy (FS) \citep{Jakobsen2022, Birkmann2022, Rigby2023} mode with the PRISM, and the Mid Infrared Instrument (MIRI) in the Low Resolution Spectroscopy (LRS) \citep{Kendrew2015, Kendrew2016, Rigby2023} mode, on 2024-06-10 and 2024-09-05 through program GO-4516 (PI  S.~W.~Jha; \citealt{JhaJWST2024}).

The data were reduced using the publicly available \textit{JWST} reduction pipeline \citep{bushouse_pipeline} for standard reduction procedures, including bias and dark subtraction, background subtraction, flat-field correction, wavelength calibration, flux calibration, rectification, outlier detection, resampling, and spectral extraction. Inspection of the 2D spectral images and the extracted 1D spectra available on the Mikulski Archive for Space Telescopes (MAST)\footnote{\url{https://mast.stsci.edu/portal/Mashup/Clients/Mast/Portal.html}} indicates that the automated reductions are of high quality and further improvements to the data reduction are not necessary.

\section{Photospheric Phase}
\label{sec:Photosepheric_phase}

\subsection{Light Curves}
\label{subsec:Lightcurves}

The photometric data for SN\,2023qov are modeled with fits to the Spectral Adaptive Light curve Template (SALT) \citep{guy_salt_2005,guy_salt2_2007}. Specifically, we use the SALT2 model, which has been shown to have better matches to pre-maximum photometry when compared to the more recently trained SALT3 \citep{Rigault_SALT2justification_2024}. We perform these fits through the \texttt{SNCosmo} package \citep{barbary_sncosmo_2016}. The SALT2 model fits SN~Ia light curves across each band with three primary parameters: $x_{o}$, $x_{1}$, and $c$. A smaller value of $x_{1}$ indicates faster evolution; $c$ describes the SN color and is parameterized as $c = (B-V)_{\rm max} - \left\langle B - V\right\rangle$.

From the SALT2 fits, we define a peak \emph{B}-band magnitude 
\begin{equation}
    m_{B} = -2.5\,\log \left( x_0 \right) + 10.5\, ,
\end{equation}
where by convention $m_B = 10.5$\,mag corresponds to $x_0 = 1$ \citep{kenworthy_salt3_2021}.

Applying the color and stretch corrections directly to the distance modulus yields 
\begin{equation}
    \mu_{\text{obs}} = m_{B} + \alpha\,x_1 - \beta\,c - M_{B}\, , \label{eq:salt_mod}
\end{equation}
where $\mu_{\text{obs}}$ represents the inferred distance modulus, and 
$\alpha$, $\beta$, and $M_{B}$ are parameters that are obtained by performing a cosmological fit with a full sample of SNe~Ia. 

We correct for Milky Way dust extinction in our SALT2 model fit by assuming a Milky Way $R_V = 3.1$ and setting $E_{\rm MW}(B-V) = 0.033$\,mag, which is the value along the line of sight obtained from the dust maps of \cite{schlegel_new_1990}, recalibrated by \cite{schlafly_measuring_2011}. As SN\,2023qov is well separated from its elliptical host (NGC 70290), we expect little host-galaxy extinction contribution, so the SALT2 color parameter ($c$) is likely representative of the intrinsic color of SN\,2023qov (as well as potential contribution from its CSM, discussed in \autoref{subsec:dust}).

\begin{figure}
    \centering
    \includegraphics[width=1\linewidth]{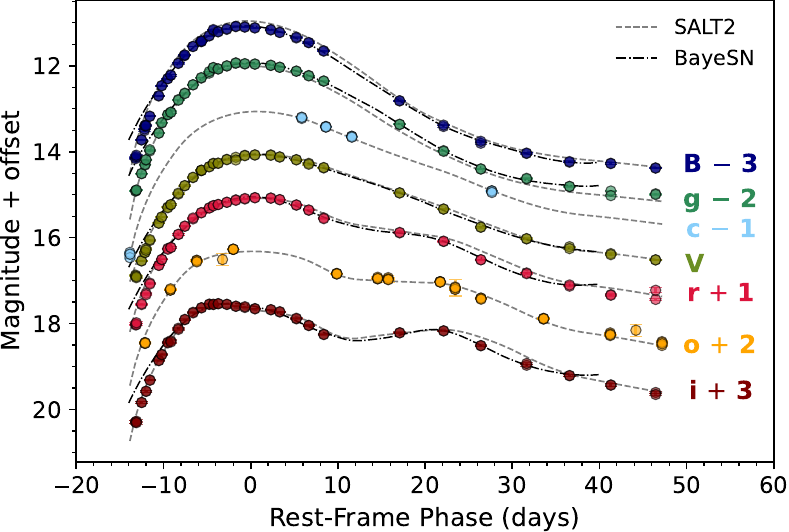}
    \caption{Early light-curve photometry fit with SALT2 and BayeSN models. SALT2 is signified by the dashed gray line, and BayeSN by the dash-dotted black line. BayeSN is cut off beyond 40~days and before $-$14~days as it is best trained to fit the peak features of the light curve, and it does better than SALT2 in that range. BayeSN does not have built-in bandpasses for ATLAS \textit{o} and \textit{c}.}
    \label{fig:models}
\end{figure}

\begin{table}
    \centering
    \footnotesize
    \begin{tabular}{l|c|c|c}
        \hline
        \hline
         & ${m_{B}}$ (peak) & ${x_{1}}$ & ${c}$ \\
          \hline
        SN 2023qov & $+13.830 \pm 0.018$ & $-1.715 \pm 0.014$ & $-0.049 \pm 0.016$ \\
        SN 1994D & $+11.704 \pm 0.027$ & $-1.627 \pm 0.025$ & $-0.069 \pm 0.024$ \\
        SN 2022aaiq & $+13.327 \pm 0.018$ & $-0.294 \pm 0.031$ & $-0.133 \pm 0.017$\\
        SN 2024gy & $+12.916 \pm 0.021$ & $+0.063 \pm 0.012$ & $+0.273 \pm 0.019$\\
        SN 2021aefx & $+11.980 \pm 0.002$ & $+ 0.292 \pm 0.009$ & $+ 0.013 \pm 0.001$ \\
    \hline
    \hline
    \end{tabular}
    \caption{SALT2 parameters for SN\,2023qov, SN\,1994D, SN\,2022aaiq, SN\,2024gy, and SN\,2021aefx.}
    \label{tab:SALT3 comparison}
\end{table}

The SALT2 fit of SN\,2023qov (\autoref{fig:models}) shows a peak \textit{B}-band magnitude of $13.830 \pm 0.018$ at $t_0 = 60193.186 \pm 0.015$ MJD. SN\,2023qov has a fast-declining light curve with $x_{1}=-1.715 \pm 0.014$, and is slightly red at $c=-0.049 \pm 0.016$. Table \ref{tab:SALT3 comparison} compares these results with those of four other regular SNe~Ia. SN\,1994D is a close match in color, decline rate, and host separation, and the others are sampled with \textit{JWST}. 

We also fit SN\,2023qov with \texttt{BayeSN}, a hierarchical Bayesian SN Ia light-curve-fitting code \citep{mandel_hierarchical_2022}. BayeSN differs from the SALT2 model in that it treats host-galaxy dust and intrinsic SN spectral energy distribution (SED) variations separately, instead of combining the two properties into a single color parameter. It thus fits for the \emph{V}-band extinction ($A_V$) directly and marginalizes over the covariances of the SN SED residuals to capture both of these physically distinct factors simultaneously. BayeSN also fits for the distance modulus directly, assuming a fiducial cosmology from the results of \cite{riess_24_2016}.

Additionally, we employ \texttt{Supernova Python} (\texttt{SNooPy}) to fit the light curve of SN\,2023qov \citep{burns_carnegie_2011,burns_carnegie_2014}. \cite{phillips_reddening-free_1999} define the traditional $\Delta m_{15}(B)$ as the decrease in $B$-band magnitude from peak brightness to $+$15 rest-frame days. We determine this to be $\Delta m_{15}(B) = 1.47 \pm 0.05\,$mag via the BayeSN $B$-band fit. SNooPy defines the $\Delta m_{15}$ parameter in their \textit{EBV Model 2},\footnote{\url{https://users.obs.carnegiescience.edu/cburns/SNooPyDocs/html/models.html}} from which we determine $\Delta m_{15} = 1.458 \pm 0.063\,$mag and the host $E(B-V) = -0.042 \pm 0.064$\,mag, consistent with our assumption of minimal host contribution. This $\Delta m_{15}$ is slightly systematically different from $\Delta m_{15}(B)$ \citep{burns_carnegie_2011}, although we note that they are very similar for SN\,2023qov. \cite{burns_carnegie_2014} introduce the color-stretch parameter $s_{BV}$, which reduces sensitivity to reddening by relating $t_0$ to the $B-V$ color-curve peak. Fitting the same \textit{EBV\_model2} with instead the $s_{BV}$ parameter, we find $s_{\rm BV} = 0.759 \pm0.032$, and again a near-zero host reddening of $E(B-V) = -0.014\pm0.063$\,mag. Using the relations between $s_{\rm BV}$, $\Delta m_{15}$, and the SALT2 $x_1$ parameter, we verify the consistency of these three results.

\texttt{SNooPy} includes a built-in bolometric light-curve routine, for which we adopt the ``direct'' method. This method integrates the observed bandpasses to estimate fluxes, then fits blackbodies to predict the IR and UV contributions. The procedure yields a pseudobolometric light curve, shown in \autoref{fig:phot} (from hereon, we succinctly refer to it as a ``bolometric" light curve). We estimate uncertainties by perturbing each photometric point according to its reported uncertainty and recomputing the bolometric light curve 500 times. This represents a total optical contribution to the bolometric luminosity, as the IR echo discussed in \autoref{subsec:dust} is not taken into account in this estimate. We use SciPy's \texttt{curve\_fit} \citep{harris_numpy_2020, scipy} to fit a line to the late-time data, which results in a predicted optical bolometric magnitude decline of $0.022 \pm0.001$\,mag\,day$^{-1}$. The peak of the light curve is fit with a spline to determine a peak optical Luminosity of $L_{\rm bol}= 4.12 \pm 0.82 \times 10^{42}$\,erg\,s$^{-1}$. We determine the uncertainty by bootstrapping 2000 times, and assume an additional systematic error of $\sim15\%$ owing to the lack of UV or NIR data. Using  Arnett's rule (\citealt{arnett_type_1982}; see Equation 5 of \citealt{Dhawan_Ni56mass_2016}) and this luminosity estimate, we predict a $^{56}$Ni mass of $M_{56} = 0.21 \pm 0.04$\,M$_{\odot}$. This is quite different from its SALT2 analogue SN\,1994D, which is predicted to have  $M_{56}\approx 0.6$--0.8\,M$_{\odot}$ \citep{hoflinch_1994D_1995,Cappellaro_1994D_1997}.

 % Fitting with the SNooPy \textit{color\_model}, employing a distribution of $R_v$ values surrounding $R_v = 0.0$ as a prior \footnote{\url{https://users.obs.carnegiescience.edu/cburns/SNooPyDocs/html/models.html\#color-model}}, we determine $s_{BV} = 0.759 \pm 0.032$, $R_v = \pm$, and $E(B-V)\_host = $.

% \begin{equation}
%     s_{BV} = 0.955 - 0.458 (\Delta m_{15} (B) - 1.1)
% \end{equation}
% and similarly to the SALT2 $x_1$ parameter:
% \begin{equation}
%     x_{1} = -0.006 + 5.98 (s_{BV} - 1) - 5.55 (s_{BV} - 1)^2 
% \end{equation}
% Plugging in, we find $s_{BV}$ from \textit{color\_model} is in agreement with the SALT2 $x_1$ parameter (Table \ref{tab:SALT3 comparison}) and the $\Delta m_{15}$ from \textit{EBV\_model}.

\subsection{Distance}
\label{subsec:distance}

We recover a distance modulus from our BayeSN fit of $\mu_{\rm BayeSN} = 32.78 \pm 0.11$\,mag, corresponding to a luminosity distance of $36.0\pm1.8$\,Mpc. As mentioned, BayeSN also fits for the selective $V$-band host dust extinction, which in this case is $A_V = 0.12\pm0.05$\,mag. \textbf{We note that this value is consistent with 0 mag to within $2.4\sigma$, which agrees with the result of \texttt{SNooPy} and our assumption of little line-of-sight dust given its isolated environment.}

The distance modulus derived from the SALT2 method in Equation~\ref{eq:salt_mod} requires assumed fit parameters for decline-rate correction $\alpha$, color correction $\beta$, and fiducial \textit{B}-band absolute magnitude $M_B$. Recently, it has been shown that the optimal decline-rate correction parameter ($\alpha$) changes depending on whether the SN~Ia is fast-declining or slow-declining \citep{Garnavich_2023,Larison_environments_2024,Ginolin_ztf_environment_2024}. Therefore, to make an accurate correction owing to the decline rate, we fit these standardization parameters using the sample from \citet{Larison_environments_2024}, with only the $z<0.06$ (to avoid Malmquist bias) and fast-evolving ($x_1 < -1$) SNe~Ia selected, using the same fiducial cosmology as our BayeSN model to ensure an accurate comparison. For this fit, we employ a Markov chain Monte Carlo (MCMC) technique, which also fits for an intrinsic scatter parameter $\sigma_{\rm int}$, which captures the variations of the SNe~Ia away from the assumed cosmology after standardization. The values we recover for our fit parameters are $\alpha = 0.167\pm0.033$, $\beta = 2.539\pm0.156$, $M_B = -19.115\pm0.064$, and $\sigma_{\rm int} = 0.156\pm0.012$. With this standardization, we find a distance modulus of $\mu_{\rm SALT2} = 32.78$\,mag, with a contribution to the uncertainty from light-curve fitting of $0.05$\,mag. With this measurement, the SALT2 and BayeSN values are in excellent agreement. Such a test between the two methods was also performed by \citet{Newman_TRGB_2025} on six regular SNe~Ia, from which they found similar agreement between the BayeSN and SALT2 methods when making similar selection cuts on the SN~Ia cosmological samples.

We note that the BayeSN fit to the light curve of SN\,2023qov has a lower reduced $\chi^2$ than the SALT2 fit (both shown in \autoref{fig:models}), with reduced $\chi^2$ values of $1.9$ and $2.7$ for the BayeSN and SALT2, fits respectively. Therefore, we recommend using the final distance recovered from BayeSN, with full statistical uncertainty as stated above.

\subsection{Colors and Velocities}

\begin{figure}
    \centering
    \includegraphics[width=\linewidth]{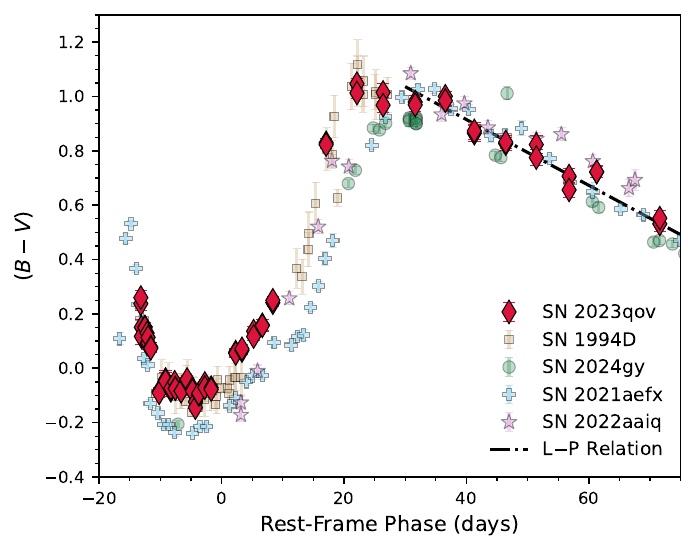}
    \caption{The $B-V$ color curve of SN\,2023qov compared with SNe~Ia 1994D \citep{Hicken_94Dphot_2009}, 2021aefx \citep{Hosseinzadeh_23bee_2023}, 2022aaiq, and 2024gy \citep{Kwok_2022aaiq24gy_2025}, corrected for Milky Way dust extinction $E_{\rm MW}(B-V)$. Rest-frame phase is relative to the time of \textit{B} maximum. The line shows the Lira-Phillips relation \citep{lira_light_1996,phillips_reddening-free_1999}, where $t(V_{\rm max}) \approx 60194.7 \pm 0.1$, or 1.5 days after $t(B_{\rm max})$.} 
    \label{fig:colorcurves}
\end{figure}

\autoref{fig:colorcurves} shows the $B-V$ color of SN\,2023qov, corrected for Milky Way extinction from the NED extinction calculator\footnote{\url{https://ned.ipac.caltech.edu/extinction_calculator}}. The color was calculated by interpolating and subtracting the \textit{B}- and \textit{V}-band magnitudes. Similar to the comparison SNe~Ia, SN\,2023qov is blue at maximum. Early emission shows little blue excess compared to SN\,2021aefx, though both show some especially defined ``infant-reddening" within the first few days. SN\,2023qov's color is similar to its SALT2 match, SN\,1994D. The \textit{B} and \textit{V} light curves have a gap at $10-20$ days for SN\,2023qov. Beyond 30 days, SN\,2023qov begins to better match the color of the comparison SNe, as it falls along the Lira–Phillips relation \citep{lira_light_1996,phillips_reddening-free_1999}.

\begin{figure}
    \centering
    \includegraphics[width=\linewidth]{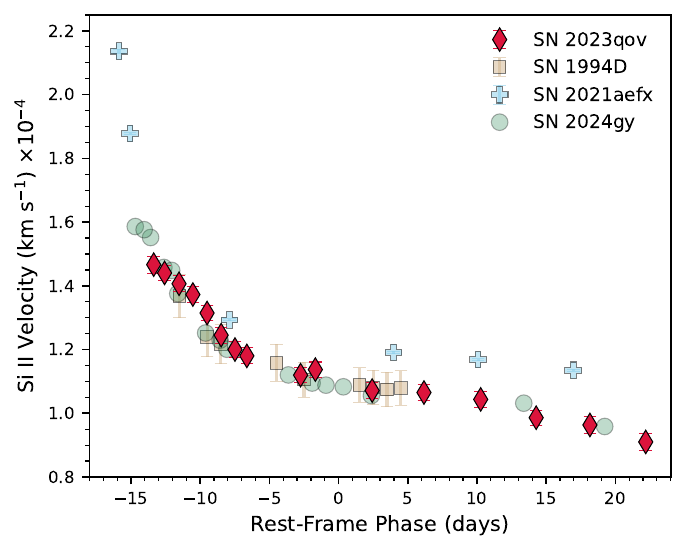}
    \caption{The \ion{Si}{2} ($\lambda_{\text{rest}}=6355$ \AA) velocity of SN\,2023qov in early phases compared against SN\,1994D \citep{Patat_SN1994Dvelocity_1996}, SN\,2021aefx \citep{Ashall_2021aefxearly_2022}, and SN\,2024gy \citep{li_sn2024gy_2024}. We do not include SN\,2022aaiq as a comparison object owing to a lack of early-time spectra.} 
    \label{fig:siII_velocity}
\end{figure}

With the spectra collected before 25 days post-explosion, we measure the \ion{Si}{2} absorption-line minimum ($\lambda_{\rm rest} = 6355$\,\AA) by fitting a single Gaussian curve with a linear component to correct for the underlying photospheric continuum with \texttt{curve\_fit} from \texttt{NumPy} \citep{harris_numpy_2020}. Adopting the resulting parameters as initial values, we use the Python package \texttt{emcee} \citep{foreman-mackey_emcee_2013} to determine the best fit and uncertainty via the Monte Carlo method. From this result we calculate the \ion{Si}{2} velocity at maximum absorption over the first 25 days. \autoref{fig:siII_velocity} compares the calculated \ion{Si}{2} velocity curve against other SNe~Ia. SN 2023qov is similar in velocity to most comparisons, \textbf{and can be put into the normal-velocity subclass according to the classification scheme proposed by \citet{Wang_2009_subclasses}}. The outlier is SN\,2021aefx, and SN\,2023qov is again similar to SN\,1994D. Both of these comparisons also fit well with strong, multicomponent, high-velocity \ion{Si}{2} models (see \citealt{silverman_siII_2015}), though the data here are single Gaussian fits, likely dominated by the photospheric velocity.

\section{Nebular Phase
\label{sec:NebularLineProfiles}}

\subsection{Dust Emission}
\label{subsec:dust}

The most striking feature of the MIR spectroscopy of SN\,2023qov is the additional flux continuum that can be seen starting at 3.5\,$\mu$m (see \autoref{fig:JWSTspec_e1_e2}), which is indicative of IR reemission of light absorbed from the SN by circumstellar dust. We do not see evidence of silicate dust or polycyclic aromatic hydrocarbons (PAHs) in the spectra \citep{draine_optical_1984,Draine_ISM_2011}, nor is there evidence of any molecules (such as CO or SiO). As an initial characterization, we fit a blackbody model to each \textit{JWST} epoch in the range 3.5--6.4\,$\mu$m which is free from strong emission lines. We bootstrap the datasets 5000 times assuming a $10\%$ uncertainty. For epoch 1 ($+276$\,d), we measure a blackbody temperature, $T_\mathrm{\rm BB, Ep1} = 476 \pm 7$\,K, and for Epoch 2 ($+363$\,d), $T_\mathrm{\rm BB, Ep2} = 397 \pm 5$\,K. Because this is not a comprehensive dust model, we expect the uncertainties are underestimated; we show example blackbody fits at $T=475$\,K and $T=400$\,K for epochs 1 and 2, respectively, in \autoref{fig:JWSTspec_e1_e2}. Nevertheless, we see clear support for a cooling temperature of $\sim 80$\,K in 90 days. Integrating the total blackbody flux we calculate a blackbody radius  $r_{\rm BB} = \sqrt{L_{\rm BB}/(4\pi \sigma T_{\rm BB}^{4})}$, given in \autoref{tab:dustmass}. Because the blackbody model assumes the dust is in an optically thick spherical shell, the resulting radii are only a lower limit to the physical size. Indeed, our calculated blackbody radii are within the expected ejecta radii at both epochs (taking $r_{\rm ejecta} \approx v_{\rm Bmax} t$ and $v_{\rm Bmax}\approx 11,000$\,km\,s$^{-1}$, results are listed in \autoref{tab:dustmass}), an unlikely region for pre-existing dust to survive. This implies that the dust is either  optically thin, not spherically distributed, or both. We do not see evidence of dust formation (e.g., through molecular lines; see \autoref{sec:Discussion}). For our subsequent analysis, we assume the dust is optically thin and at a larger physical radius.

\begin{figure*}
    \centering
    \includegraphics[width=\linewidth]{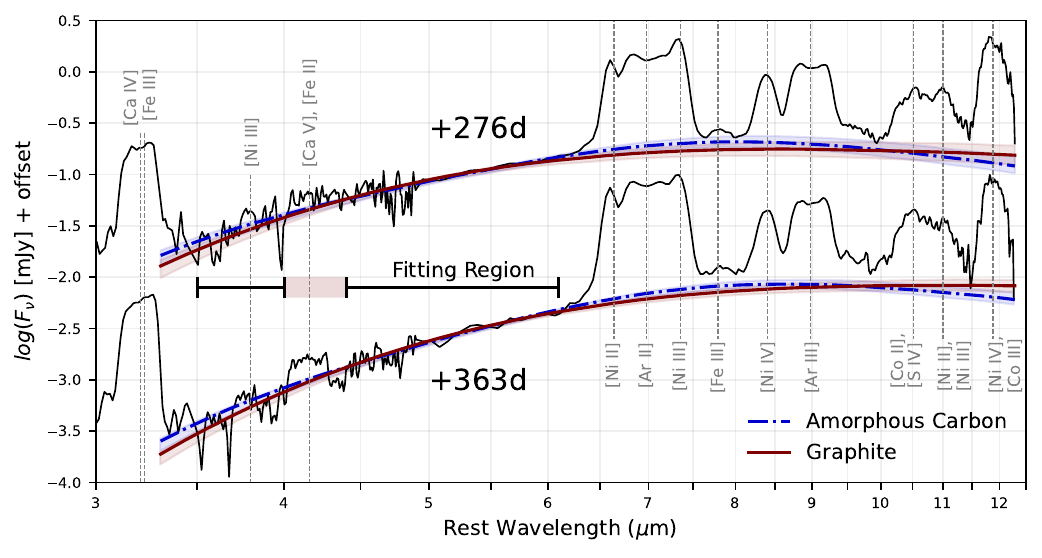}
    \caption{Two epochs of \textit{JWST} spectra fit against carbonaceous dust models from \cite{Jager_AMCdust_1996} (dash-dotted line) and \cite{draine_optical_1984} (solid line), both with $2\sigma$ model uncertainties overplotted. The region 3.5--6~$\mu$m was used to fit the models, with the blended [\ion{Ca}{5}]/[\ion{Fe}{2}] $4.2\,\mu$m feature cut out (denoted with the red shaded region).}
    \label{fig:dust_fits}
\end{figure*}

To better estimate the dust mass and temperature giving rise to this MIR continuum, we fit two optically thin carbonaceous dust models, shown in \autoref{fig:dust_fits}. We fit a similar region to the blackbody fits, cutting out the $4.21\,\mu$m [\ion{Ca}{5}]/[\ion{Fe}{2}] feature. To determine uncertainties, we bootstrap the datasets 5000 times for each model and epoch. In order to calculate flux as a function of wavelength, we employ an optically thin dust-emission model,

\begin{equation}
    F_{\nu} = \frac{\kappa M}{d_{SN}^2}B_{\lambda}(\lambda,T)\, .
\end{equation}
% $B_{\lambda}(\lambda,T)$
We take $d_{\rm SN} \approx 36.0$\,Mpc (\autoref{subsec:dust}), and $B_{\lambda}(\lambda,T)$ is the Planck blackbody function. The final parameter is the opacity,

\begin{equation}
    \kappa = \frac{3}{4\rho a} Q_{\rm abs}(\lambda)\, ,
\end{equation}

\noindent where $Q_{\rm abs}$ is the absorption efficiency, $\rho$ is the density of the grain material, and $a$ is the average grain radius. We apply the optical graphite absorption efficiency model from \cite{draine_optical_1984}, assuming a density of 2.24\,g\,cm$^{-3}$ and an average grain size $0.1\,\mu$m, representative of standard astronomical graphitic dust and commonly adopted in light-echo and circumstellar-dust calculations \citep{Draine_2007}. The amorphous carbon (AMC) model from \cite{Jager_AMCdust_1996} is also used, giving a model for $Q_{\rm abs}/a$, and we assume a density of 1.85\,g\,cm$^{-3}$ \citep{Rouleau_AMCdensit_1991,Zubko_AMCdensity_1996}. Varying these dust properties has an effect of a factor of a few on the measured dust masses, so the masses should be regarded as an order-of-magnitude estimate. The exact dust parameters do not substantially change our findings. The resulting parameters are shown in \autoref{tab:dustmass}. Both models give consistent, or somewhat decreasing, dust masses across the two epochs.

% The graphite model seems to fit the overall shape of the dust continua across the two epochs better than the amorphous carbon.
\begin{table*}[]
    \centering\begin{tabular}{c|c|c|c|c|c|c|c}
    \hline
    \hline
        Epoch & $r_{\rm ejecta}$ [lyr]& $T_{\rm BB}$ [K] & $r_{\rm BB}$ [lyr] & $M_{\rm Gra}$ [M$_\odot~$]& $T_{\rm Gra}$ [K] & $M_{\rm AMC}$ [M$_\odot~$] & $T_{\rm AMC}$ [K]\\
        \hline
        $+276$\,d & $\sim 3 \times10^{-2}$ &$476 \pm 7$& $(7.1 \pm 0.2) \times 10^{-3}$&$3.51^{+0.71}_{-0.60}\times10^{-4} $ & $375\pm9$ & $2.09 ^{+0.39}_{-0.32}\times10^{-4}~$& $419 \pm 10$\\
        $+363$\,d & $\sim 4\times10^{-2}$ &$397\pm 5$& $(6.5 \pm 0.2) \times 10^{-3}$ &$3.09^{+0.41}_{-0.34}\times10^{-4}$ & $325 \pm 5$ & $1.66^{+0.20}_{-0.19}\times10^{-4}$ & $362\pm6$\\
        \hline
    \end{tabular}
    \caption{Ejecta radii and dust parameters for the blackbody, graphite, and amorphous carbon (AMC) models over both nebular epochs.}
    \label{tab:dustmass}
\end{table*}

%While we assume a blackbody fit for the dust emission, and this fit is a good upon visual inspection, the dust may not be an empirical blackbody.  Furthermore, the blackbody emission tells us nothing of the composition of the dust, and so this assumed fit cannot be used to search for progenitor tracers, such as molecular grains which would point to recently formed dust.  

\subsection{\textbf{Thin} Spherical Shell IR Echo Model}
\label{subsec:echo}

\begin{figure}
    \centering
    \includegraphics[width=1\linewidth]{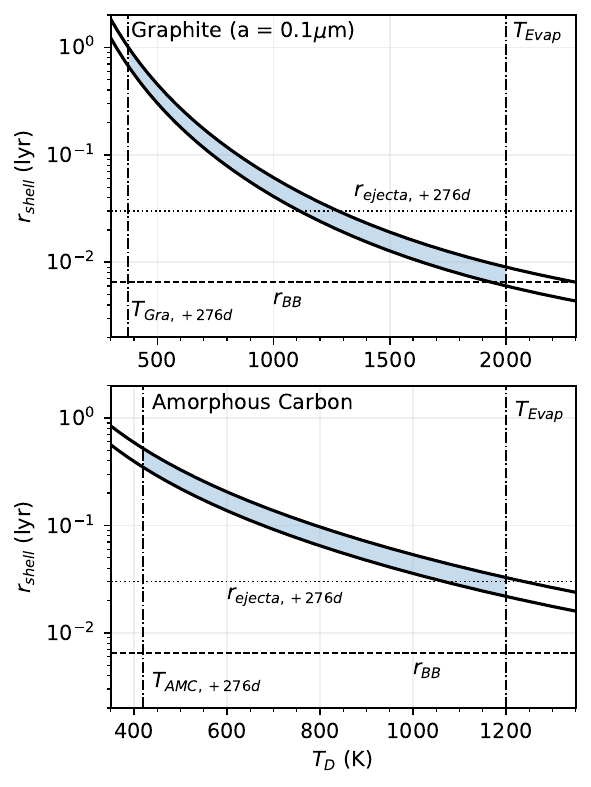}
    \caption{Possible shell radii ($r_{\rm shell}$) of CSM around SN\,2023qov, plotted against the maximum continuum temperature to which the dust is heated $T_{\rm D}$, from a spherical dust IR echo model (\autoref{subsec:echo}). Here, $T_{\rm Gra/AMC,+276d}$ is our measured temperature for both dust species (\autoref{subsec:dust}), corresponding to a lower limit on $T_{\rm D}$, and $T_{\rm Evap}$ is the maximum temperature allowed for the species of dust. The blackbody radii are $r_{\rm BB}$, determined in \autoref{subsec:dust}. We plot the radii of both species for a range of 2$\sigma$ of $L_{\rm bol}$,  determined in \autoref{subsec:Lightcurves}, to account for the uncertainty in the true optical bolometric luminosity of SN\,2023qov.}
    \label{fig:dust_radii}
\end{figure}

To further constrain its properties, we model the dust emission as IR reemission of the optical radiation from the SN, assuming the majority of initial dust heating occurred from peak brightness. We define $L_{\rm bol}$ as our calculated peak luminosity in the UV and optical (\autoref{fig:phot}). Following a derivation similar to that of \cite[][see their Section 3.3]{fox_disentangling_2010}, we \textbf{assume a thin shell, and} balance the radiation absorbed and reemitted by the dust, \textbf{setting} $L_{\rm abs} = L_{\rm D}$, \textbf{and} resulting in an effective shell radius

\begin{equation}
    r_{\rm shell}^2 = \frac{3}{64} \frac{L_{\rm bol}}{a \rho \sigma T_{\rm SN}^4} \frac{\int B_{\nu}(T_{\rm SN})Q_{\rm abs}(\nu) d\nu}{\int B_{\nu}(T_{\rm D})\kappa (\nu) d\nu}\, .
\end{equation}
\noindent
Here $T_{\rm SN}$ is the effective temperature of the SN at peak brightness, which we estimate as $T_{\rm SN} \approx 10,500$\,K by fitting a continuum to the peak optical spectrum. The result does not depend precisely on $T_{\rm SN}$. Also, $T_{\rm D}$ is the maximum temperature to which the dust is heated. We take this to be the temperature at the first epoch for both dust compositions, though this is a lower limit, as the dust temperature may have exceeded this at an earlier epoch than observed. We also note an upper limit of temperatures, at $T_{\rm Evap, ~Gra} \approx 2000$\,K and $T_{\rm Evap, ~AMC} \approx 1200$\,K based on estimates of their sublimation temperatures \citep{Draine_ISM_2011,Blasius_AMCsublimation_2012}. \autoref{fig:dust_radii} shows the possible radii given these constraints.

We find the upper limits on shell radii of $R_{\rm Gra} < 0.85 \pm 0.22\,$lyr, and $R_{\rm AMC} < 0.44 \pm 0.11$\,lyr. Uncertainties were estimated by propagating the errors of $T_{\rm D}$ and $L_{\rm bol}$. These results are consistent with the timescale expected of an IR echo owing to CSM within $1$\,lyr of the SN. Declining dust emission is expected to begin once the light from the peak of the SN has passed through the bulk of the dust. Thus, since we see a decline in dust emission sometime $\lesssim 1$\,yr, we expect the majority of 
dust to be within or near $1$\,lyr for most geometries (see \citealt{Maeda_dustHJK_2015} and \citealt{fox_disentangling_2010} for further discussion on the relationship between phase and geometry in IR echos). Higher-cadence MIR sampling of a ``dusty'' Type Ia SNe like SN\,2023qov would allow for the further constraining of dust heating and cooling timescales, as well as their corresponding distances and geometries. 

\subsection{Emission-Line Morphology}
\label{subsec:nebularlines}

To identify nebular emission lines, we follow the predictions of \cite{Blondin_nebular_2023} and the analyses of other normal SNe~Ia SN\,2021aefx \citep{Kwok_2021aefxJWST_2023, Ashall_21aefxJWST_2024}, SN\,2022aaiq, and SN\,2024gy \citep{Kwok_2022aaiq24gy_2025}. SN\,2023qov shares the same dominant spectral lines with these objects, but  there are a few differences. We identify the $7.788\,\mu$m feature to be [\ion{Fe}{3}] \citep{Blondin_nebular_2023}, which is more clearly present than in other analyses. The flat-topped $4.21\,\mu$m feature, only strongly present in the second epoch, appears to be a blended combination of primarily [\ion{Ca}{5}] and a possible weak contribution from two distinct [\ion{Fe}{2}] lines, though low S/N in this region makes characterization difficult. The blended feature in the $\sim 10.5\,\mu$m region is somewhat distinct in morphology from previous objects, consistent with being blended [\ion{S}{4}], [\ion{Co}{2}], [\ion{Ni}{2}], and [\ion{Ni}{3}] lines, with the [\ion{S}{4}] being especially pronounced.

Following the procedures described by \cite{Kwok_2021aefxJWST_2023, Kwok_22pul_2024,Kwok_2022aaiq24gy_2025}, we subtract the best-fit blackbody models and fit NIR and MIR emission-line profiles for both epochs. In doing this, we assume the dust is pre-existing and thus not impacting the line profiles.  We select the dominant emission lines in the 6--12.4\,$\mu$m region to fit: [\ion{Ni}{2}]~6.63, 10.679\,$\mu$m, [\ion{Ar}{2}]~6.985\,$\mu$m, [\ion{Ni}{3}]~7.349, 11.002\,$\mu$m, [\ion{Ni}{4}]~8.405, 11.723\,$\mu$m, [\ion{Ar}{3}]~8.991\,$\mu$m, [\ion{Co}{2}]~10.521\,$\mu$m, [\ion{S}{4}]~10.510\,$\mu$m, and [\ion{Co}{3}]~11.888\,$\mu$m. These are consistent with SN~Ia abundance models \citep[e.g.,][]{Nomoto_abundancemodeling_1984,Thielemann:1986,fink_DD_2010,pakmor_violentmerger_2012,Seitenzahl_DDT_2013} and were also examined in the nebular \textit{JWST} analyses of other normal SNe~Ia. The lines are fit together in order to preserve the relative line strengths between transitions of the same ion from \cite{Blondin_nebular_2023}. The region 6--9.4\,$\mu$m is weighted more heavily in the fit in order to better constrain the blended $\sim 11\,\mu$m region. This is done via SciPy's curve$\_$fit \textit{sigma} argument, where the blended region is assumed to have double the uncertainty compared to the unblended region. This also helps to ensure the higher S/N blueward region of the MIRI spectrum is more strongly influencing the fit.

\begin{figure*}
    \centering
    \includegraphics[width=1\linewidth]{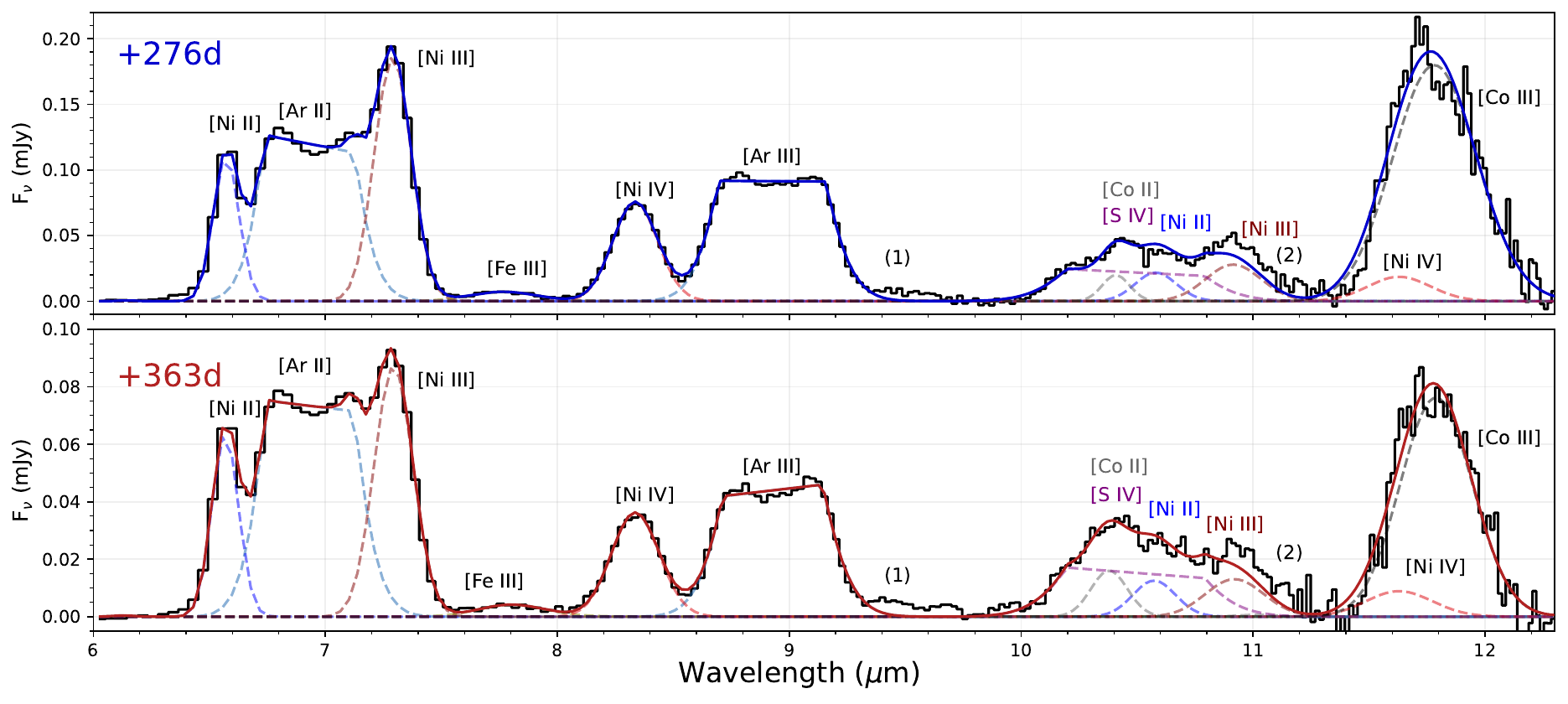}

    \caption{Nebular line fits to the blackbody-subtracted \textit{JWST}/MIRI spectra (black). The top panel shows epoch 1 ($+276$\,d); the bottom panel shows epoch 2 ($+363$\,d). The less blended features on the left side of the plot are weighted more heavily in the fit, so the lines in the heavily blended 10--11\,$\mu$m feature are well constrained, with the exemption of [\ion{S}{4}] and [\ion{Co}{2}], which do not appear outside of the blended region. Label (1) shows an ambiguous feature with low S/N, which could be various IGEs. Label (2) highlights an ambiguous excess that we attribute to either noise, an underpredicted model relative line strength of [\ion{Ni}{2}] \citep{Blondin_nebular_2023}, possible $11.24\,\mu$m [\ion{Ni}{5}] emission, or some combination of unidentified weak lines. The features are discussed in \autoref{subsec:nebularlines}.} 
    \label{fig:JWSTlines}
\end{figure*}

We fit Gaussian profiles to the Ni and Co lines, while we fit flat-topped profiles with Gaussian wings to the Ar and S lines. The shell model produces an emission line having a boxy, flat-topped shape with Gaussian wings. We primarily select these profiles based on the work of \citet{Kwok_22pul_2024}, who showed that the line profiles of these IMEs differ from the IGEs that dominate at nebular times, which also aligns with expectations from models involving detonations \citep[e.g.,][]{Blondin_nebular_2023}. The fits to the dominant lines are listed in \autoref{tab:linefits} and shown in \autoref{fig:JWSTlines}.

A notable feature in SN\,2023qov is the double-peaked (``horned'') nature of the Ar lines (and likely S, though line blending makes the shape uncertain), which are distinct from SN\,2021aefx, SN\,2022aaiq, and SN\,2024gy. This deviation from the ``flat-top'' resulting from a spherical shell geometry indicates that the emission is somewhat toroidal. \cite{Gerardy_MIR_2007} also found double-peaked, ring-like Ar emission in {\it Spitzer} MIR spectra of SN\,2005df and SN\,2003dh, which appear more pronounced than in SN\,2023qov.

The toroidal geometry still indicates  stratified ejecta with the IMEs located in outer layers, supported by our measured line widths (see \autoref{tab:linefits}). The IGEs have narrower widths than the IMEs, indicating that they are located closer to the center of the ejecta. 

\begin{deluxetable*}{rcc || r || cr}
    \tablecaption{MIR line identification and fit parameters for SN\,2023qov (\autoref{fig:dust_fits}).  The velocities are in units of $10^3$\,km\,s$^{-1}$.  In both profiles, $v_{\mathrm{off}}$ represents the offset of the central line profile from rest in velocity space. In the shell profile,  $v_{\mathrm{inner}}$ is a proxy for the width of the ``flat-top'' and $v_c$ is a proxy for the slope of the ``flat-top," the velocity offset center of the shell to the center of emission, with positive (negative) values having positive (negative) slope.  \label{tab:linefits}}
    \tablehead{
        \colhead{$\lambda_{\mathrm{rest}}$ ($\mu$m)} & \colhead{Line} & \colhead{Profile} & \colhead{Best-Fit Parameters} & \colhead{Best-Fit Parameters}
    } 
    \startdata
         \multicolumn{3}{c}{} & \multicolumn{1}{c}{\textbf{Epoch 1} ($+$276\,d)} & \multicolumn{1}{c}{\textbf{Epoch 2} ($+$363\,d)} \\
        \hline
        6.636, 10.679 & [Ni II] & Gaussian & 
        $v_{\mathrm{FWHM}}$: 5.91 $\pm$ 0.65, $v_{\mathrm{off}}^{b}$: $-$2.80 $\pm$ 0.26& 
        $v_{\mathrm{FWHM}}$: 6.12 $\pm$ 0.55, $v_{\mathrm{off}}$: $-$2.99 $\pm$ 0.24 \\
        6.985 & [Ar II] & Shell & 
        $v_{\mathrm{FWHM}}$: 20.93 $\pm$ 0.99, $v_{\mathrm{off}}$: $-$2.48 $\pm$ 0.31 &
        $v_{\mathrm{FWHM}}$: 21.09 $\pm$ 0.96, $v_{\mathrm{off}}$: $-$2.32 $\pm$ 0.25 \\
        & & & 
        $v_{\mathrm{inner}}$: 8.39 $\pm$ 0.23, $v_c$: $-$0.20 $\pm$ 0.14 &
        $v_{\mathrm{inner}}$: 8.19 $\pm$ 0.25, $v_c$: $-$0.10 $\pm$ 0.13 \\
        7.349, 11.002 & [Ni III] & Gaussian & 
        $v_{\mathrm{FWHM}}$: 7.72 $\pm$ 0.43, $v_{\mathrm{off}}$: $-$2.32 $\pm$ 0.21 &
        $v_{\mathrm{FWHM}}$: 8.12 $\pm$ 0.43, $v_{\mathrm{off}}$: $-$2.19 $\pm$ 0.24 \\
        8.405, 11.723 & [Ni IV] & Gaussian &
        $v_{\mathrm{FWHM}}$: 8.20 $\pm$ 0.61, $v_{\mathrm{off}}$: $-$2.42 $\pm$ 0.24 &
        $v_{\mathrm{FWHM}}$: 8.71 $\pm$ 0.70, $v_{\mathrm{off}}$: $-$2.42 $\pm$ 0.27 \\
        8.991 & [Ar III] & Shell &
        $v_{\mathrm{FWHM}}$: 19.73 $\pm$ 0.67, $v_{\mathrm{off}}$: $-$2.20 $\pm$ 0.17 &
        $v_{\mathrm{FWHM}}$: 18.68 $\pm$ 0.77, $v_{\mathrm{off}}$: $-$1.94 $\pm$ 0.19 \\
        & & &
        $v_{\mathrm{inner}}$: 7.80 $\pm$ 0.27, $v_c$: $-$0.01 $\pm$ 0.15 &
        $v_{\mathrm{inner}}$: 7.25 $\pm$ 0.34, $v_c$: 0.17 $\pm$ 0.16 \\
        10.510 & [S IV] & Shell &
        $v_{\mathrm{FWHM}}$: 19.44 $\pm$ 0.85, $v_{\mathrm{off}}$: $-$1.00$^{a}$ &
        $v_{\mathrm{FWHM}}$: 21.26 $\pm$ 0.91, $v_{\mathrm{off}}$: $-$1.00$^{a}$\\
        & & &
        $v_{\mathrm{inner}}$: 8.50$^{a}$, $v_c$: $-$0.50$^{a}$&
        $v_{\mathrm{inner}}$: 8.50$^{a}$, $v_c$: $-$0.50$^{a}$\\
        10.521 & [Co II] & Gaussian &
        $v_{\mathrm{FWHM}}$: 4.00 $\pm$ 2.65, $v_{\mathrm{off}}$: $-$3.21 $\pm$ 0.73 &
        $v_{\mathrm{FWHM}}$: 5.34 $\pm$ 1.94, $v_{\mathrm{off}}$: $-$4.04 $\pm$ 0.41 \\
        11.888 & [Co III] & Gaussian &
        $v_{\mathrm{FWHM}}$: 10.88 $\pm$ 0.19, $v_{\mathrm{off}}$: $-$2.63 $\pm$ 0.08 &
        $v_{\mathrm{FWHM}}$: 9.32 $\pm$ 0.22, $v_{\mathrm{off}}$: $-$2.45 $\pm$ 0.09 \\
        \hline
    \enddata
    \tablenotetext{a}{[\ion{S}{4}] $v_{\mathrm{FWHM}}$ and $v_{\rm inner}$ are fixed based on the unblended [\ion{Ar}{2}] and [\ion{Ar}{3}] features; owing to heavy blending, [\ion{S}{4}] and [\ion{Co}{2}] are poorly constrained.}
    \tablenotetext{b}{The systematic offset in $v_{\mathrm{off}}$ for each line is indicative of an off-center explosion, with the epicenter of the explosion being closer to the observer along the line of sight relative to the rest frame.}
\end{deluxetable*}

\begin{figure*}
    \centering
    \includegraphics[width=.8\linewidth]{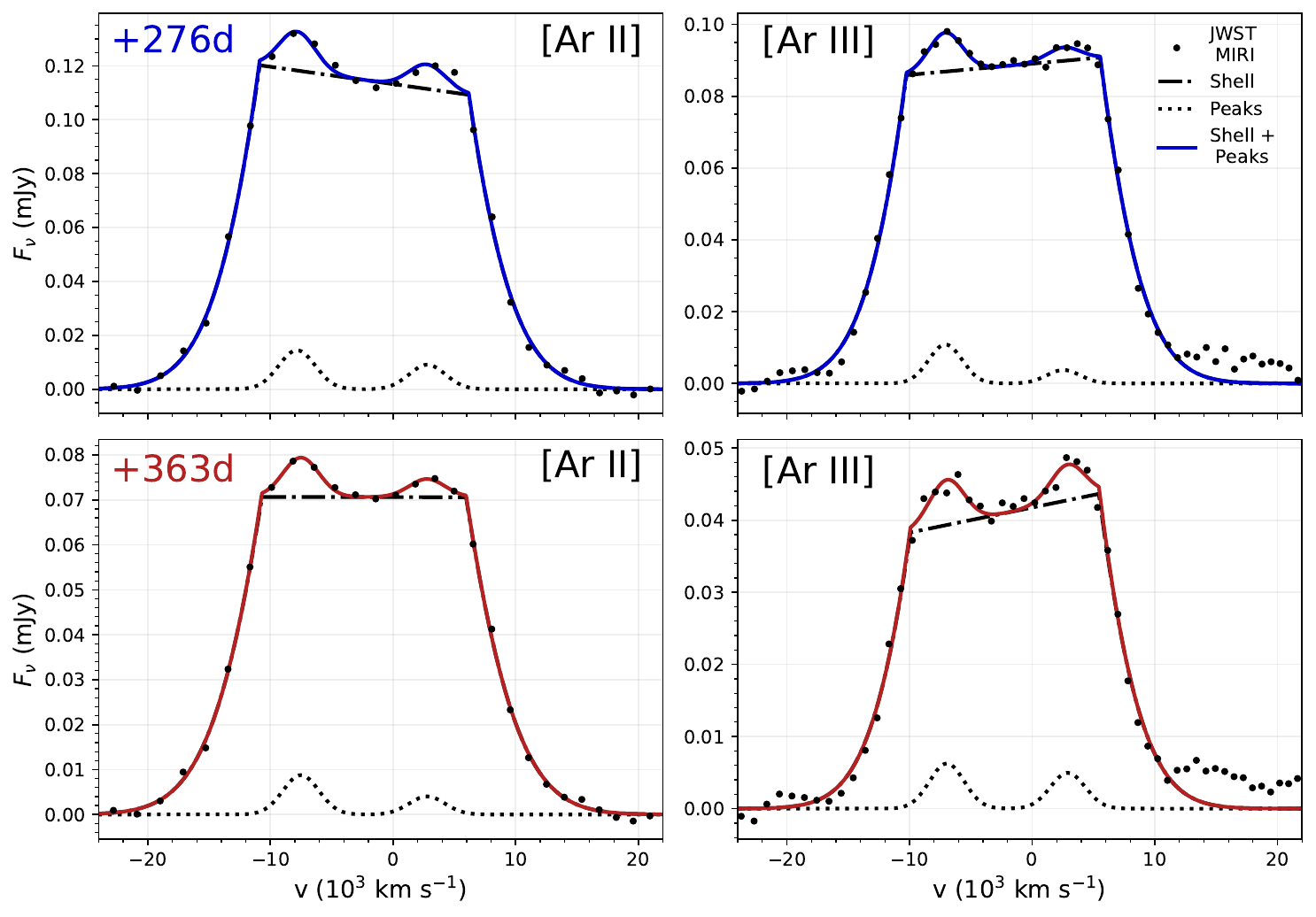}
    \caption{Simple torus model fit of [\ion{Ar}{2}] and [\ion{Ar}{3}] features marked in \autoref{fig:JWSTlines}. Upper panels are from the first epoch ($+276$\,d) and lower are from the second ($+363$\,d) of \textit{JWST} MIRI data of SN\,2023qov. The peaks and shell are combined into one model, with the Gaussian contributions constrained to 5--10\% of the amplitude and full width at half-maximum intensity (FWHM) of the shell to constrain degeneracies.}
    \label{fig:torus}
\end{figure*}

In the overall fit (\autoref{fig:JWSTlines}) we employ a shell model for the Ar emission, but to further analyze the structure of these horned line profiles, we fit a shell with two smaller Gaussian contributions to represent the torus, shown in \autoref{fig:torus}. To isolate the Ar lines, we first subtract the neighboring Ni line contributions using the previous fits. We then combine our slanted shell model with two smaller Gaussian peaks. We estimate and constrain the flux amplitude of the Gaussian contributions to be 5--10\% of the amplitude of the shell. This is physically motivated, as the bulk geometry is expected to be shell-like, while the additional Gaussian components are smaller density variations. Due to the degeneracies between the model parameters, the data could support other solutions with the ratio between shell and Gaussian contribution varying more significantly, but we find the constrained solution more physically plausible. The parameters of the constrained fit are given in \autoref{tab:torus}. As with previous fits, uncertainties were determined via bootstrap.

\begin{table}
\centering
\small
\setlength{\tabcolsep}{9pt}
\renewcommand{\arraystretch}{1}
\makebox[\columnwidth][c]{}
\begin{tabular}{l||c||c}
\multicolumn{3}{c}{\textbf{Epoch 1} ($+276$ d)} \\
\hline
\hline
Ion & [\ion{Ar}{2}] & [\ion{Ar}{3}] \\
\hline

\multicolumn{3}{l}{\textit{Shell Component}} \\
\hline
$A_{\mathrm{shell}}$ & $0.37 \pm 0.01$ & $0.28 \pm 0.02$ \\
$v_{\mathrm{FWHM}}$ & $21.23 \pm 0.17$ & $19.80 \pm 0.24$ \\
$v_{\mathrm{off}}$ & $-2.53 \pm 0.04$ & $-2.26 \pm 0.05$ \\
$v_{\mathrm{in}}$ & $8.52 \pm 0.07$ & $7.90 \pm 0.10$ \\
$v_c$ & $-0.17 \pm 0.04$ & $0.09 \pm 0.08$ \\
\hline

\multicolumn{3}{l}{\textit{Gaussian Components}} \\
\hline
$A_{\mathrm{blue}}$ & $0.015 \pm 0.002$ & $0.011 \pm 0.003$ \\
$A_{\mathrm{red}} / A_{\mathrm{blue}}$ & $0.63 \pm 0.24$ & $0.35 \pm 0.32$ \\
$v_{\mathrm{FWHM}}$ & $3.50^{a}$ & $3.20 \pm 0.73$ \\
$\Delta v$ & $10.62 \pm 0.36$ & $9.63 \pm 0.67$ \\
\hline
\end{tabular}

\vspace{0.4em}

\begin{tabular}{l||c||c}
\multicolumn{3}{c}{\textbf{Epoch 2} ($+363$ d)} \\
\hline
\hline
Species & [\ion{Ar}{2}] & [\ion{Ar}{3}] \\
\hline

\multicolumn{3}{l}{\textit{Shell Component}} \\
\hline
$A_{\mathrm{shell}}$ & $0.20 \pm 0.01$ & $0.13 \pm 0.01$ \\
$v_{\mathrm{FWHM}}$ & $21.51 \pm 0.14$ & $19.06 \pm 0.33$ \\
$v_{\mathrm{off}}$ & $-2.37 \pm 0.03$ & $-2.03 \pm 0.08$ \\
$v_{\mathrm{in}}$ & $8.32 \pm 0.06$ & $7.68 \pm 0.15$ \\
$v_c$ & $-0.002 \pm 0.054$ & $0.22 \pm 0.04$ \\
\hline

\multicolumn{3}{l}{\textit{Gaussian Components}} \\
\hline
$A_{\mathrm{blue}}$ & $0.009 \pm 0.002$ & $0.006 \pm 0.002$ \\
$A_{\mathrm{red}} / A_{\mathrm{blue}}$ & $0.46 \pm 0.17$ & $0.79 \pm 0.29$ \\
$v_{\mathrm{FWHM}}$ & $3.50^{a}$ & $3.33 \pm 0.65$ \\
$\Delta v$ & $10.28 \pm 0.33$ & $9.86 \pm 0.74$ \\
\hline
\end{tabular}

\vspace{0.6em}

{\footnotesize
$^{a}$Values held to upper limits.
}

\caption{Parameters from combined shell and double Gaussian fits (\autoref{fig:torus}). Here, $\Delta v$ is the separation between peaks; $A_{\mathrm{red}}/A_{\mathrm{blue}}$ is the red-to-blue Gaussian amplitude ratio. Shell-component parameters are introduced in \autoref{tab:linefits}, but are refit here with the simple torus model. Velocities and widths are in units of $10^3$\,km\,s$^{-1}$, while amplitudes are given in units of mJy.}
\label{tab:torus}
\end{table}

This model shows that the line profile corresponds well to a bulk shell-like structure, with overdensities along the line of sight, that combine to a torus-like structure. Measurements of the shell thickness and the separation and widths of the added Gaussian components indicate that the toroidal component extends over the \emph{inner} $\sim 25$\% of the shell and lies near the line of sight on opposite sides of the ejecta relative to the observer. Further modeling may determine how inclination and different ejecta density profiles affect this toroidal line profile. We discuss possible implications of these line profiles in \autoref{sec:Discussion}.

% In the case of fitting the lines of [\ion{Ni}{2}]~6.63~$\mu$m, [\ion{Ar}{2}]~6.985~$\mu$m, and [\ion{Ni}{3}]~7.349$\mu$m, in Epoch 2, the model resulted in bimodal distributions for several model parameters, regardless of prior selection.  Visual inspection of several fits in the bootstrapping routine showed an over-assumption of the [\ion{Ar}{2}]~6.985~$\mu$m line.  To resolve this, we limited our selection to just fits which overlapped the best-fit model.  For this line set, we increased our total number of perturbed spectra so as to have at least 1,000 measurements for the uncertainty histograms.

\section{Discussion and Conclusions}
\label{sec:Discussion}

We have presented early- to late-time ($-13$\,days to $1$\,yr) photometric and spectroscopic observations of SN\,2023qov, a nearby, normal SN~Ia. Its early-time color and \ion{Si}{2} velocity evolution are similar to those of SN\,1994D, another fast-declining normal SN~Ia. Using our well-sampled photometry, we measure a distance of $36.0 \pm 1.8$\,Mpc, using BayeSN. We present ground-based optical spectra of SN\,2023qov spanning 1\,yr. Additionally, two epochs (+276 and +363\,d) of \textit{JWST} NIR+MIR spectra are analyzed, with which we make the first robust detection of a dust continuum in a normal SN~Ia. We further fit the nebular line profiles of SN\,2023qov, and find that it is consistent with other \textit{JWST}-observed SNe~Ia in showing stratified asymmetric shells of IMEs surrounding Gaussian-like IGEs. 

The cooling continuum emission detected in the \textit{JWST} observations of SN\,2023qov seems to indicate an IR echo of circumstellar dust heated by radiation from the SN. We do not find clear evidence for dust being freshly formed after the explosion. The direct detection of dust in the MIR spectrum of SN\,2023qov is the first such discovery in a normal SN~Ia, but previous direct detections of dust in subtypes of SNe~Ia have been seen in the SN Ia-CSM 2018evt \citep{Wang_CSMDust_2024} and the 03fg-like SN\,2022pul \citep{Siebert:2024,Kwok_22pul_2024}. After this discovery in SN\,2023qov, reinspection of the \textit{JWST} spectrum of the normal Type Ia SN\,2021aefx presented by \cite{Kwok_2021aefxJWST_2023} reveals potential evidence for a weak dust continuum (see their Figure 1). The existence of dust in SN\,2023qov, SN Ia-CSM 2018evt, SN\,2022pul, and perhaps SN\,2021aefx, conflicts with long-held SN~Ia theory of a nearly pristine CSM \citep{Gerardy_MIR_2007, Gomez_dust_2007, GomezClark_Remnantdust_2012}. In the case where dust is pre-existing, local SN~Ia environments are not as pristine as previously thought, and in the case where the dust is formed in the SN explosion, SNe~Ia may contribute to the production of interstellar dust along with core-collapse SNe and pre-explosion stellar sources, though there is currently little evidence to support the post-explosion scenario.

Fitting the MIR dust continua at the two \textit{JWST} epochs with various dust models (\autoref{subsec:dust}) confirms that the dust is best fit by blackbody-like carbonaceous dust models, and we measure dust masses on the order of $M_{\rm dust} \approx 10^{-4}$\,M$_{\odot}$. In the $\sim$80~days between the two epochs, the dust mass does not change substantially. We do not detect tracers of dust creation, such as carbon monoxide or other molecules often seen in core-collapse SNe \citep[e.g.,][]{spyromilio_carbon_1996,spyromilio_carbon_2001,rho_dust_2009,park_23ixf_2025}, or an identifiable second warm dust component \citep[e.g.,][]{fox_disentangling_2010}. We also cannot identify a significant rebrightening in the late-time optical light curves, as has been seen in some SN~Ia-CSM \citep[e.g.,][]{Graham_PTF11kx_2017,Tsalapatas_2020aeuh_2025}. Thus, the dynamical interaction with the ejecta has either not yet occurred, was missed in phases without optical photometry, or is weak.

Our estimated IR light-echo distance (\autoref{fig:dust_radii}) suggests that the dust is within $1$\,lyr of the SN. This alone does not prove that the dust is from the progenitor system, but considering the remote location and elliptical host, little ``environmental'' dust is expected. Higher cadence MIR follow-up observations of ``dusty'' SNe~Ia such as SN\,2023qov would allow the shell radii, geometry, and composition to be further constrained, with implications for the origin of dust in and around SNe~Ia.\textbf{ }We summarize the possible physical implications of the dust under each origin scenario below.

\begin{itemize}
    \item \emph{Formed by the Progenitor}

If the dust is pre-existing and involved with the progenitor system, future models must account for the presence of pre-existing carbonaceous circumstellar dust. Material may be left over from pre-explosion classical nova outbursts (as is seen in cataclysmic variable stars; 
\citealt{Evans_novae_1990}), or winds and unstable mass transfer with some evolved donor star \citep{Nomoto_SDCSM_2018}. These cases are naturally associated with single-degenerate or core-degenerate progenitor scenarios, while comparable mechanisms for producing circumstellar dust in double-degenerate systems are less well established \citep{Pakmor_2022_DD}. The presence of pre-existing dust constrains the nature of the progenitor, namely that whatever system preceded SN\,2023qov underwent mass loss or outbursts.

    \item \emph{Formed by the Supernova}

Despite the lack of emission from common dust tracers like molecular CO in our observations, the formation of interstellar dust in the ejecta cannot be completely ruled out. Other work (e.g., \citealt{Kumar_Iax_2026}) has already shown that low-luminosity thermonuclear SNe are likely capable of hosting dust formation within hundreds of days. Upon inspection of the late-time optical spectra of the SN (\autoref{fig:JWSTspec_e1_e2}), it is not clear whether extinction on the red side of the emission features is present, as would be expected if the dust were formed in the explosion. If the dust detected in SN\,2023qov  was formed during the explosion, it is worth considering that at least some normal SNe~Ia contribute to interstellar dust populations. MIR sampling is required at earlier epochs of other ``dusty" SNe~Ia, to detect or rule out formation tracers such as CO.

    \item \emph{Environmental Dust Sheet}
    
Previous SN~Ia light echoes that likely occurred in purely environmental dust sheets are SN\,1991T \citep{Sparks_91Tlightecho_1999,schmidt_1994_1991T}, SN\,1998bu \citep{Cappellaro_98bulightecho_2001}, SN\,2006X \citep{Crotts_06Xlightecho_2008,Wang_SN2006X_2008}, and SN\,2014J \citep{Crotts_14Jlightecho_2015,Foley_SN2014J_2014,Yang_SN2014J_2018}. These events occurred in comparatively dusty, star-forming environments. SN\,2023qov is in a remote environment (\autoref{fig:image}), in which we do not see \ion{Na}{1} absorption indicative of a dust-rich local environment, and only exhibits slight reddening (\autoref{fig:colorcurves}). These environmental differences, together with the modeling presented in \autoref{subsec:echo}, favor a circumstellar origin for the dust. If the dust were as far away as an interstellar dust sheet from the SN, it is less likely the SN could have heated it to our measured temperatures. However, owing to the lack of strong constraints on the evolution and geometry of the light echo, we cannot entirely rule out the case of an interstellar dust sheet. If this echo arose from environmental dust, it would suggest that even remote SNe~Ia may require nonnegligible host-extinction corrections.

\end{itemize} 

The strong detection of dust in SN\,2023qov (as compared with other normal SNe~Ia) leads to the consideration that dust may have some effect on SN light curves, in particular the color variation, and thus the standardization of normal SNe~Ia. Dust is expected to redden and extinguish optical light, and SN\,2023qov has a slightly red light curve a few days post-explosion; however, the extent to which the early light curve is reddened by possibly pre-existing dust is unclear. \cite{Bulla_2018_reddening} determined that the reddening is quite sensitive to the dust parameters, which we can now infer via MIR spectra. Using this to analyze reddening of a wider population will allow for robust characterization of the color variation caused by dust. This will expand physical interpretations for the empirical color corrections made for cosmological analyses. It may further clarify the existing degeneracies between line-of-sight host extinction, pre-existing circumstellar dust emission and extinction, and the intrinsic color of the SN. A concrete determination of the cosmological implications of dust in SNe~Ia will require the continued expansion of the sample of normal SN~Ia observed by \textit{JWST} in the MIR.

In addition to the dust emission, SN\,2023qov shows strong stable nickel emission in the NIR and MIR nebular phases. The light curves of normal SNe~Ia are primarily powered by the radioactive decay of $^{56}$Ni into $^{56}$Co and at later times from $^{56}$Co into $^{56}$Fe. At nebular times, all $^{56}$Ni will have already decayed, so nebular Ni lines trace stable Ni ($^{58}$Ni, $^{60}$Ni, etc.). The stable Ni lines (of which $^{58}$Ni is the most abundant isotope) in SN\,2023qov have a Gaussian profile, consistent with a central distribution of stable Ni. When compared with other objects in the \textit{JWST} sample of SNe~Ia, SN\,2023qov appears to lack the multicomponent broad base plus narrow top \ion{Ni}{2} features which are strong in SN\,2022aaiq and SN\,2024gy \citep{Kwok_2022aaiq24gy_2025}. They more closely align with the Gaussian Ni profiles of the 91bg-like \citep{Filippenko_91bg_1992b, Leibundgut_91bg_1993} SN\,2022xkq \citep{Derkacy_22xkq_2024}. Further, the nickel lines are weaker in amplitude and widths relative to Ar when compared with these normal SN~Ia counterparts, being somewhere between them and SN\,2022xkq. The formation of this quantity and the distribution of $^{58}$Ni in SN\,2022aaiq and SN\,2024gy may require either the amount of high-density burning present in a near-Chandrasekhar-mass white dwarf \citep{Blondin_nickel_2022, blondin_subluminous_2018} or higher metallicities and mixing in a sub-Chandrasekhar-mass white dwarf \citep{Shen_subchandraNickel_2018,Flors_subchandranebular_2020}. The narrower, nearly Gaussian Ni line profiles in SN\,2023qov show continued variation in Ni distribution between SNe~Ia, which may imply differences in explosion mechanisms. Further MIR sampling of the nickel features of SNe~Ia in quiescent environments will help to clarify whether they have a distinct explosion mechanism (i.e., a double detonation) compared with those in star-forming environments. For a more in-depth overview of the implications of stable nickel in SNe~Ia, see \cite{Blondin_nickel_2022}, \cite{Kumar_nickel_2026}, and \cite{Kwok_2022aaiq24gy_2025}.

In the fits to the NIR and MIR emission lines, we see continued preference for a flat-topped model for the Ar and S lines, similar to SN\,2021aefx \citep{Kwok_2021aefxJWST_2023}, SN\,2024gy, and SN\,2022aaiq \citep{Kwok_2022aaiq24gy_2025}. While the fit of a shell model, ideal for the Ar and S lines in these objects, provides a better fit for SN\,2023qov than a Gaussian, the profile is double-peaked instead of purely flat-topped (\autoref{fig:torus}). This implies that the emission may arise from an asymmetric ring, rather than a spherical shell of emission (or some combination thereof). \citet{Prust_torus_2025} showed that a similar geometry can arise from companion interaction, with a conical cavity ``carved out'' by the companion as the ejecta expand. It has not yet been shown, though, that this scenario can fully recreate the toroidal geometry and velocity structure in the IMEs, while having the IGEs continue to appear Gaussian. Regardless, the broad flat-tops of the Ar and S lines --- in comparison with the narrower Gaussian-shaped Ni and Co lines --- indicate  stratified ejecta, with IMEs located outside an IGE-dominated core. 

The emission-line morphology and the presence of dust in SN\,2023qov could hint at the cause of the variances in evolutionary characteristics among normal SNe~Ia. Asymmetries in the ejecta can cause viewing-angle-dependent changes to the light curve, as well as the possibility of reddening due to dust. Concrete conclusions linking photometric properties, such as color, decline rate, and brightness, to NIR and MIR spectral characteristics, including these line asymmetries and the presence of circumstellar dust, will require a larger sample of high-quality late-time \textit{JWST} observations of normal SNe~Ia.

\section*{Acknowledgements}

This work is based in part on observations made with the NASA/ESA/CSA {\it James Webb Space Telescope}. The data were obtained from the Mikulski Archive for Space Telescopes at 
the Space Telescope Science Institute (STScI), which is operated by the Association of Universities for Research in Astronomy (AURA), Inc., 
under NASA contract NAS5-03127 for {\it JWST}. These observations are associated with {\it JWST} program GO-4516 (PI S.~W.~Jha). The specific observations analyzed can be accessed via \dataset[doi: 10.17909/6h34-tq75]{https://doi.org/10.17909/6h34-tq75}.

The SALT data presented here were obtained with Rutgers University programs 2023-1-MLT-008, 2024-1-MLT-003 (PI S.~W.~Jha), and 2024-1-MLT-004 (PI L.~Kwok). We thank Rudi Kuhn, Antoine Mahoro, Moses Mogotsi, and Lee Townsend for making these observations.
This work makes use of data from the Las Cumbres Observatory network. The LCO team is supported by NSF grant AST-2308113. 
Based in part on data acquired at the ANU 2.3-metre telescope. The automation of the telescope was made possible through an initial grant provided by the Centre of Gravitational Astrophysics and the Research School of Astronomy and Astrophysics at the Australian National University and through a grant provided by the Australian Research Council through LE230100063. We acknowledge the traditional custodians of the land on which the telescope stands, the Gamilaraay people, and pay our respects to elders past and present.
We also used observations collected at the European Southern Observatory under ESO programme 114.27JL.001, and the European Organisation for Astronomical Research in the Southern Hemisphere under ESO programme 114.27JL. This work was supported by the ``Action Thématique de Physique Stellaire'' (ATPS) of CNRS/INSU PN Astro cofunded by CEA and CNES.
Some of the data presented herein were obtained at the W. M. Keck Observatory, which is operated as a scientific partnership among the California Institute of Technology, the University of California, and NASA; the observatory was made possible by the generous financial support of the W. M. Keck Foundation.

C.W.M. acknowledges support by the National Science Foundation Graduate Research Fellowship Program under grant 2444108. \textit{JWST} and ground-based supernova research at Rutgers University is supported by awards JWST-GO-02072 and NSF AST-2407567, respectively. S.W.J. is also grateful for support from a Guggenheim Fellowship. C.L. acknowledges funding through \textit{JWST} program grants GO-06541, GO-06585, and GO-05324. Support for M.D. was provided by Schmidt Sciences, LLC. L.A.K. is supported by NASA through Hubble Fellowship grant HF2-51579.001-A awarded by STScI, which is operated by AURA, Inc. under contract NAS5-26555 for NASA.

Time-domain research by the University of Arizona team and D.J.S. is supported by NSF grants AST-2308181, 2407566, and 2432036. B.B. received support from the Hungarian National Research, Development and Innovation Office grants OTKA PD-147091, the János Bolyai Research Scholarship of the Hungarian Academy of Science and from the HUN-REN Hungarian Research Network.  C.B. acknowledges support from Chandra Theory grants TM0-21004X and TM1-22004X, XRISM Guest Scientist grant 80NSSC23K0634, and NSF grants AST-2307865 and 2511539. S.H. thanks the LSST-DA Data Science Fellowship Program, which is funded by LSST-DA, the Brinson Foundation, the WoodNext Foundation, and the Research Corporation for Science Advancement Foundation; her participation in the program has benefited this work. A.J. acknowledges support by the Swedish Research Council (grant 2018–03799). 

K. Maeda was supported by Japan Society for the Promotion of Science (JSPS) KAKENHI grants JP24KK0070 and JP24H01810. K. Maguire acknowledges funding from Horizon Europe ERC grant 101125877.  M.R.S. is supported by an STScI Postdoctoral Fellowship. T.T. acknowledges support from NSF grant AST-2205314 and NASA ADAP award 80NSSC23K1130. J.H.T. acknowledges Horizon Europe ERC grant 101125877. M.S. acknowledges financial support provided under the National Post Doctoral Fellowship (N-PDF; File Number PDF/2023/002244) by the Science \& Engineering Research Board (SERB), Anusandhan National Research Foundation (ANRF), Government of India. J.V. is supported by Hungarian NKFIH-OTKA grant K142534. L.Z.W. is sponsored by the National Natural Science Foundation of China (NSFC) grant 12573050, the Chinese Academy of Sciences South America Center for Astronomy (CASSACA) Key Research Project E52H540301, and the Chinese Academy of Sciences (CAS) through a grant to the CASSACA. The work of X.W. is supported by  %National Natural Science Foundation of China 
NSFC grants 12288102 and 12033003, the Ma Huateng Foundation, and the New Cornerstone Science Foundation through the XPLORER PRIZE.

A.V.F.’s research group at UC Berkeley acknowledges financial assistance from the Christopher R. Redlich  
Fund, Gary and Cynthia Bengier, Clark and Sharon Winslow, Alan Eustace and Kathy Kwan (W.Z. is a Bengier-Winslow- 
Eustace Specialist in Astronomy),       
Timothy and Melissa Draper, Briggs and Kathleen Wood,  Ellyn and Alan Seelenfreund (T.G.B. is Draper-Wood-Seelenfreund Specialist in Astronomy), and numerous other donors.

\facilities{ANU (WiFeS), ATLAS, \textit{JWST} (NIRSpec/MIRI), Keck (LRIS), LCO/GSP (FTS/FLOYDS), SALT (RSS), VLT (XShooter).}

\software{Astropy \citep{astropy_i,astropy_ii,astropy_iii}, BayeSN \citep{mandel_hierarchical_2022}, Matplotlib \citep{hunter_matplotlib}, \hyperlink{https://github.com/mmechtley/ned_extinction_calc}{NED Extinction Calculator}, Numpy \citep{numpy}, Pandas \citep{reback_pandas}, PyRAF \citep{pyraf_2012}, PySALT \citep{Crawford_pysalt_2010}, jwst \citep{bushouse_pipeline}, Scipy \citep{scipy}, SNCosmo \citep{barbary_sncosmo_2016}, SNooPy \citep{burns_carnegie_2011}. Jupyter notebooks used to complete this work are available upon request.}

%% For this sample we use BibTeX plus aasjournals.bst to generate the
%% the bibliography. The ample631.bib file was populated from ADS. To
%% get the citations to show in the compiled file do the following:
%%
%% pdflatex sample631.tex
%% bibtext sample631
%% pdflatex sample631.tex
%% pdflatex sample631.tex

\bibliography{sn2023qov}{}
\bibliographystyle{aasjournal}

%% This command is needed to show the entire author+affiliation list when
%% the collaboration and author truncation commands are used.  It has to
%% go at the end of the manuscript.
%\allauthors

%% Include this line if you are using the \added, \replaced, \deleted
%% commands to see a summary list of all changes at the end of the article.
%\listofchanges

\end{document}